\documentclass[11pt,a4paper]{article}
\usepackage{jheppub}
\usepackage{bbm}
\usepackage{youngtab}
\usepackage{rotfloat}
\usepackage{stmaryrd}
\usepackage{amsfonts,amssymb,amsmath,multirow}
\usepackage{mathtools}
\usepackage{arydshln}
\usepackage{blkarray, bigstrut}
\usepackage{tikz-cd}
\usepackage{caption}
\usepackage[all,cmtip]{xy}
\usepackage{fancybox}
\usepackage{xcolor}

\newtheorem{conj}{Conjecture}
\newtheorem{qn}{Question}

\usepackage{bm} 

\usepackage[T1]{fontenc}

\DeclareSymbolFont{Letters} {U}{zeur}{m}{n}
\DeclareMathSymbol{\talpha}  {\mathalpha}{Letters}{"0B}
\DeclareMathSymbol{\tbeta}   {\mathalpha}{Letters}{"0C}
\DeclareMathSymbol{\tgamma}  {\mathalpha}{Letters}{"0D}

\newcommand{\bea}{\begin{eqnarray}}
\newcommand{\eea}{\end{eqnarray}}
\newcommand{\be}{\begin{equation}}
\newcommand{\ee}{\end{equation}}

\newcommand{\Z}{{\mathbb Z}}
\newcommand{\R}{{\mathbb R}}
\newcommand{\C}{{\mathbb C}}

\def\Tr{{\rm Tr \,}}

\def\frak{\mathfrak}

\def\tilde{\widetilde}
\def\hat{\widehat}
\def\bar{\overline}

\def\CF{{\mathcal F}}

\def\CH{{\mathcal H}}
\def\CI{{\mathcal I}}

\def\CK{{\mathcal K}}
\def\CL{{\mathcal L}}
\def\CM{{\mathcal M}}
\def\CN{{\mathcal N}}
\def\CO{{\mathcal O}}

\def\CR{{\mathcal R}}

\def\CT{{\mathcal T}}
\def\CU{{\mathcal U}}

\def\CW{{\mathcal W}}

\newcommand{\cp}{{\mathbb{C}}{\mathbf{P}}}

\renewcommand{\bar}{\overline}
\renewcommand{\hat}{\widehat}

\DeclareMathAlphabet      {\mathbi}{OML}{cmm}{b}{it}

\definecolor{ao(english)}{rgb}{0.0, 0.5, 0.0}

\usepackage{version}

\excludeversion{NB}

\preprint{CALT-2020-019}

\title{Rozansky-Witten geometry of Coulomb branches and logarithmic knot invariants}

\author[1]{Sergei Gukov}
\author[1]{Po-Shen Hsin}
\author[2]{Hiraku Nakajima}
\author[3]{Sunghyuk Park}
\author[4]{Du Pei}
\author[1]{Nikita Sopenko}
\date{\today}
\affiliation[1]{Walter Burke Institute for Theoretical Physics, California Institute of Technology,\\ Pasadena, CA 91125, USA}
\affiliation[2]{Kavli Institute for the Physics and Mathematics of the Universe (WPI), The University of Tokyo, 5-1-5 Kashiwanoha, Kashiwa, Chiba, 277-8583, Japan}
\affiliation[3]{Department of Mathematics, California Institute of Technology, Pasadena, CA 91125, USA}
\affiliation[4]{Center of Mathematical Sciences and Applications, Harvard University, Cambridge, MA 02138, USA}

\abstract{By studying Rozansky-Witten theory with non-compact target spaces we find new connections with knot invariants whose physical interpretation was not known. This opens up several new avenues, which include a new formulation of $q$-series invariants of 3-manifolds in terms of affine Grassmannians and a generalization of Akutsu-Deguchi-Ohtsuki knot invariants.}

\begin{document}
\setcounter{tocdepth}{2}
\maketitle

\section{Introduction}

In 1996, Rozansky and Witten \cite{Rozansky:1996bq} proposed a novel way of constructing 3-manifold invariants, given a choice of a hyper-K\"ahler manifold $X$. As pointed out in its subsequent mathematical formulations \cite{MR1671737,MR1671725}, the space $X$ only needs to be a holomorphic symplectic manifold. Although this generalization will not play a major role in this paper, in part because our main examples of $X$ will come from Coulomb branches of 3d $\CN=4$ theories, it remains true that the Rozansky-Witten theory provides an intriguing link between geometry and topology, namely the geometry of hyper-K\"ahler manifolds and topology of 3-manifolds:
\be
X \qquad \xleftrightarrow[~~~~]{~~~~} \qquad \text{3d~ ``TQFT''}
\label{XTQFT}
\ee

Another intriguing aspect of the Rozansky-Witten theory, responsible for much of our motivation, is that it is not quite a TQFT in general. This is why we use the quotes on the right-hand side of \eqref{XTQFT}. As pointed out already in \cite{Rozansky:1996bq}, when $X$ is non-compact, the cutting and gluing properties of the resulting topological invariants in general do not obey the standard axioms of a TQFT, in the sense of Atiyah~\cite{MR1078014}. On the other hand, Coulomb branches of non-trivial 3d $\CN=4$ theories are always non-compact! 
So, how does one reconcile these facts?
Of course, it does not necessarily mean that the Rozansky-Witten invariants do not exist when $X$ is non-compact, although their definition certainly becomes more subtle. On the contrary, we wish to illustrate here that more interesting topological invariants of knots and 3-manifolds are to be found precisely when $X$ is non-compact.

Therefore, in this paper we specifically focus on {\it non-compact} examples of $X$.
One of our main goals is to develop the dictionary between geometry and topology \eqref{XTQFT} for non-compact $X$. In other words, we wish to develop practical tools for computing the Rozansky-Witten invariants $Z_{\text{RW} [X]} (M_3)$ for a fairly general non-compact target space $X$ and for a general 3-manifold $M_3$. As part of this challenge, we need to understand the cutting and gluing formulae in the Rozansky-Witten theory with non-compact target space $X$. What makes this task non-trivial and interesting is that the space of states in cohomology of the topological supercharge $Q$ on a genus-$g$ surface is \cite{Rozansky:1996bq}:
\be
\CH (\Sigma_g) \; = \;
\bigoplus_{q=0}^{\dim_{\C} X}
H_{\bar \partial}^q (X,(\wedge^* V)^{\otimes g})
\; = \;
\begin{cases}
\bigoplus\limits_{l=0}^{2n} H^{0,l} (X) \,, & g=0 \quad (\Sigma_g = S^2) \\
\bigoplus\limits_{l,m=0}^{2n} H^{l,m} (X) \,, & g=1 \quad (\Sigma_g = T^2) \\
\vdots & 
\end{cases}
\label{HRW}
\ee
This space is well-defined when $X$ is compact, but as pointed out earlier Coulomb branches of 3d $\CN=4$ theories are never compact. If $X$ is non-compact, the space $\CH (\Sigma_g)$ becomes infinite-dimensional and leads to additional difficulties, that range from summing over an infinite set of states in cutting and gluing formulae to the very definition of $\CH (\Sigma_g)$ itself that requires extra care and depends on the asymptotic behavior of the metric on $X$.

Related to this is another puzzle, that persists even when $X$ is compact. The cutting and gluing relations in a genuine 3d TQFT are encoded \cite{MR2654259} in the algebraic data of a modular tensor category (MTC) that, among other things, includes the data of $S$ and $T$ matrices which furnish a representation of the modular group $SL(2,\Z)$. This category is what a 3d TQFT should assign to a circle $S^1$. On the other hand, mathematical formulations of the Rozansky-Witten invariants mentioned earlier suggest that the relevant category is (a variant of) the derived category of the coherent sheaves on $X$, see {\it e.g.} \cite{MR2661534}.
While this category fits in well with the answer for the space of states \eqref{HRW} it does not carry, for general $X$, all the rich structure of a modular tensor category. A natural question, then, is what plays the role of MTC in a Rozansky-Witten theory with non-compact target space $X$.
\begin{qn}
	What algebraic structure --- that, suggestively, we call $\text{MTC} [X]$ --- underlies the cutting and gluing (up to codimension-2) in Rozansky-Witten theory with target $X$?
\label{qn:MTC}
\end{qn}
There are some hints that the answer to this question may be quite simple and interesting. The first hint comes from the simplest example of a non-compact hyper-K\"ahler manifold, $X=\C^2$. This space (or, its ALF version $X = \text{Taub-NUT}$) is the Coulomb branch of 3d Seiberg-Witten gauge theory~\cite{Seiberg:1996nz}, so that the Rozansky-Witten invariants in this case are equal to the Seiberg-Witten invariants of $M_3$. The latter depend on the choice of additional structure, namely the choice of Spin$^c$ structure on $M_3$, which itself has a non-trivial behavior under cutting and gluing \cite{Gukov:2019mnk}. However, apart from keeping track of this additional structure, the Reidemeister-Turaev-Milnor torsion of $M_3$ computed by the Seiberg-Witten theory \cite{MR1418579,MR1666856} has a very simple behavior under cutting and gluing: it is basically multiplicative \cite{MR528965,MR1968575}. In other words, it requires no infinite sums over the states in \eqref{HRW} and effectively behaves as if the Hilbert spaces were one-dimensional.

This somewhat surprising behavior is in good agreement with another hint, that comes from the study \cite{Gukov:2015sna, Dedushenko:2018bpp} of the Rozansky-Witten invariants for $X = \CM_H (G,C)$, the moduli space of $G$-Higgs bundles on a curve $C$ (that has nothing to do with $M_3$). In that class of models, it was argued that the answer to Question~\ref{qn:MTC} is controlled by the geometry of the fixed point set under a circle action on $X$ (known as the ``Hitchin action'' \cite{MR887284} in the literature on Higgs bundles):
\be
\text{fixed point set on }X
\quad \xleftrightarrow[~~~]{~~~} \quad
\text{MTC} [X]
\quad \xleftrightarrow[~~~]{~~~} \quad
\text{Bethe vacua}
\label{XMTCBethe}
\ee
Even though $X=\C^2$ and $X = \text{Taub-NUT}$ cannot be easily realized as moduli spaces of Higgs bundles, they enjoy a $U(1)_t$ symmetry analogous to the Hitchin action that was used in \cite{Gukov:2016gkn} and will be discussed in more detail below. The fact that this $U(1)_t$ symmetry has only one isolated fixed point on $X$ can be viewed as a reason for the relatively simple multiplicative behavior of the corresponding Rozansky-Witten invariants mentioned earlier.
Similarly, we will use the same principle \eqref{XMTCBethe} as a window into cutting and gluing properties of Rozansky-Witten invariants based on general Coulomb branches of 3d $\CN=4$ gauge theories.

Another motivation for the present work comes from parallel developments in topology, where various new invariants of knots and 3-manifolds were proposed.
This includes, for example, certain generalizations of the Alexander polynomial \cite{MR1164114,MR2466562,MR3286896} that come from the (restricted) quantum group $\bar \CU_q (\frak{sl}_2)$ at the $2p$-th root of unity, $q = e^{\frac{\pi i}{p}}$. The representation theory of this quantum group\footnote{In comparison, the famous Witten-Reshetikhin-Turaev (WRT) invariants at the same root of unity are described by a semi-simplification of this non-semisimple modular tensor category. The $S$ and $T$ matrices of a non-semisimple MTC and its semisimplification have different size, $3p-1$ and $p-1$, respectively.
} is governed by a non-semisimple MTC, in the sense of Lyubashenko \cite{MR1354257,MR1324034}, that also describes the representation theory of a logarithmic conformal field theory (log-CFT), the so-called $(1,p)$ {\it triplet} model \cite{Feigin:2005zx}.\footnote{See \cite{2018arXiv181202277N} and references therein for discussions on relations between two representation theories.} The corresponding knot invariants are often called logarithmic knot invariants.
They admit many closely related variants. For example, the singlet sector of a triplet model, naturally called $(1,p)$ {\it singlet} model, also gives rise to a non-semisimple MTC. It leads to logarithmic knot invariants that correspond to the restricted ``unrolled'' quantum group $\bar \CU_q^H (\frak{sl}_2)$ at the $2p$-th root of unity \cite{MR2441954,MR2480500,MR3795642}.

A seemingly different class of new invariants, introduced in \cite{Gukov:2016gkn,Gukov:2017kmk}, associates a $q$-series with integer coefficients to a 3-manifold $M_3$ equipped with a choice of an abelian flat connection (or, more precisely, a choice of Spin$^c$ structure \cite{Gukov:2019mnk}). These $q$-series invariants, denoted $\hat Z_b (M_3)$, provide a non-perturbative completion to all-loop perturbative expansions \cite{Dimofte:2009yn} in $SL(2,\C)$ Chern-Simons theory that behave well under cutting and gluing (surgery) operations. As such, they can (and should) be thought of as $\CU_q (\frak{sl}_2)$ invariants of $M_3$ at generic values of the parameter $|q|<1$. This perspective, in fact, was the original motivation for constructing $\hat Z_b (M_3)$ invariants, recently substantiated in \cite{SPark}.

To the best of our knowledge, Rozansky-Witten invariants were never mentioned in the previous work on logarithmic knot invariants. Similarly, neither of these were directly related to $\hat Z$-invariants (although connections between $\hat Z_b (M_3)$ and logarithmic vertex algebras were found in \cite{Cheng:2018vpl}). In this paper, not only do we want to advocate that all of these invariants are ``of the same type,'' but we also propose a web of explicit relations among them. Schematically,
\begin{equation}
\begin{tikzcd}
& {\text{Rozansky-Witten} \atop \text{invariants}} \ar[dl, leftrightarrow] \ar[dr, leftrightarrow] &  
\\
\hat Z \text{-invariants} \ar[rr, leftrightarrow] & ~~~~ & {\text{logarithmic knot} \atop \text{invariants}}
\end{tikzcd}
\label{theweb}
\end{equation}
Indeed, much like Rozansky-Witten invariants based on non-compact targets, $\hat Z$-invariants and logarithmic invariants enjoy cutting and gluing relations, but not quite of a traditional TQFT type. Another aspect they have in common is a presence of extra structures, such as Spin$^c$ structure on $M_3$ mentioned earlier.
Thus, one of our proposals is that the extra data called ``coloring'' $\omega$ in \cite{MR3286896} should be viewed as a Spin$^c$ structure, as ``$b$'' of $\hat Z_b (M_3)$.

The rest of this paper is organized as follows.
As a warm-up, in section~\ref{sec:Verlinde} we present various ways to compute $Z_{\text{RW} [X]} (M_3)$ for non-compact target spaces $X$ and 3-manifolds of the form $M_3 = S^1 \times \Sigma_g$ (or circle bundles over $\Sigma_g$), when 3d ``TQFT'' reduces to an effective 2d TQFT. In particular, motivated by Question~\ref{qn:MTC}, we focus on the algebraic structure of cutting and gluing relations, encoded in the ``quantum dimensions'' $\text{qdim} (\lambda) = S_{0 \lambda}$ and the eigenvalues of the $T$-matrix.
In section~\ref{sec:general}, we use these lessons to make a proposal for $Z_{\text{RW} [X]} (M_3)$ on general 3-manifolds, in the spirit of \cite{Dedushenko:2018bpp}. Building on these ideas, in section~\ref{sec:Zhat} we flesh out the web of relations~\eqref{theweb} with some details.

\section{Equivariant Verlinde formula for Coulomb branches}
\label{sec:Verlinde}

In a genuine 3d TQFT that obeys Atiyah's axioms \cite{MR1078014}, the invariant (``partition function'') on $M_3 = S^1 \times \Sigma_g$ computes the (super-)trace\footnote{{\it i.e.} the graded dimension} over the space of states $\CH (\Sigma_g)$ that TQFT assigns to $\Sigma_g$. Moreover, in a 3d TQFT associated with a semisimple modular tensor category, this graded dimension of $\CH (\Sigma_g)$ can be easily computed using the modular data, namely the matrix elements of the $S$-matrix \cite{Verlinde:1988sn}:
\be
Z (S^1 \times \Sigma_g)
\; = \; \text{sdim} \CH (\Sigma_g)
\; = \; \sum_{\lambda} (S_{0\lambda})^{2-2g}
\label{Verlindefla}
\ee
We will be interested in a version of this formula for {\it infinite-dimensional} spaces $\CH (\Sigma_g)$ that, nevertheless, can be expressed as a {\it finite} sum.

Namely, if the space of states $\CH (\Sigma_g)$ is infinite-dimensional, but comes equipped with a $\Z$-grading, such that each graded component is finite-dimensional, then the character of $\CH (\Sigma_g)$ is well defined. We will be interested in a generalization of the Verlinde formula \eqref{Verlindefla},
\be
\sum_n t^n \text{sdim} \CH_n (\Sigma_g)
\; = \; \sum_{\lambda} (S_{0\lambda})^{2-2g}
\label{VerlS}
\ee
such that the sum on the right-hand side runs over a finite set of ``states'' $\lambda$, and $S_{0 \lambda}$ are $t$-dependent, which for special values\footnote{While in the semisimple setting it is clear that, due to the Ocneanu rigidity, this can only happen at special values of $t$, in the non-semisimple case the Ocneanu rigidity can fail, see {\it e.g.} \cite{MR1891233,GHS}, and the entire family may potentially be related to a family of non-semisimple MTCs parametrized by $t$. (We thank Pavel Etingof for useful discussions on this point.)} of $t$ may become $S$-matrix elements of the familiar tensor categories, as it happens in various examples considered in \cite{Gukov:2015sna,Gukov:2016lki,Fredrickson:2017yka,Dedushenko:2018bpp}. In general, such specialization may require a $\zeta$-function regularization of a potentially divergent sum.

In the context of the Rozansky-Witten theory with non-compact target space $X$ this scenario is naturally realized when $X$ enjoys a holomorphic circle action with compact fixed point set.
The role of this symmetry for $Z_{\text{RW} [X]} (M_3)$ depends on whether it is tri-holomorphic or merely holomorphic. Let us start with the latter. A prototypical example of such symmetry is the Hitchin action on $X = \CM_H (G,C)$ mentioned around \eqref{XMTCBethe}. Motivated by this class of examples we shall denote a symmetry of this type by $U(1)_t$.
The corresponding equivariant parameter, $t$, is the holonomy of the background $U(1)_t$ connection along the $S^1$ in $M_3 = S^1 \times \Sigma_g$.
Following the arguments in \cite{Gukov:2015sna}, it is easy to see that, in the physical 3d $\CN=4$ theory, $t$ is the (exponentiated) mass parameter that preserves only $\CN=2$ subalgebra of the 3d $\CN=4$ supersymmetry algebra. As a result, with a non-trivial background $t \ne 1$, the theory can only be defined on 3-manifolds of the form $M_3 = S^1 \times \Sigma_g$ or, more generally, on Seifert 3-manifolds~\cite{Closset:2018ghr}.

In contrast, when the symmetry in question is tri-holomorphic, its background is compatible with the topological twist on an arbitrary 3-manifold. To distinguish this type of symmetry from $U(1)_t$, we denote\footnote{Our notations differ from \cite{Gukov:2016gkn}, where $U(1)_x$ was denoted by $U(1)_q$ and the corresponding fugacity by $q$. Here, $q$ will be reserved to denote the parameter of the quantum group.} it by $U(1)_x$, with background holonomies $x \in \CM_{\text{flat}} \big( U(1)_{\C}, M_3 \big)$. To summarize, we generally follow the notations
\begin{eqnarray}
U(1)_x: & \quad & \text{tri-holomorphic} \label{Uones} \\
U(1)_t: & \quad & \text{holomorphic, but not tri-holomorphic} \nonumber
\end{eqnarray}
generalizing them in an obvious way when larger symmetry groups are present.
On a 3-manifolds of the form $M_3 = S^1 \times \Sigma_g$, either of these symmetries can be used to compute \eqref{VerlS}, as long as the fixed point set is compact. All we need is to find an explicit expression for $S_{0 \lambda}$, as functions of the equivariant parameters $t$ and $x$. One way to do this is to recall \cite{Rozansky:1996bq} that, on $M_3 = S^1 \times S^2$, the invariant $Z_{\text{RW} [X]} (M_3)$ computes the index of the Spin$^c$ Dirac operator on $X$, which in the presence of symmetries can be evaluated using the equivariant localization. When combined with \eqref{VerlS}, it gives
\be
Z_{\text{RW} [X]} \left( S^1 \times S^2 \right)
\; = \; \int_X \widehat{A} (TX)
\; = \; \sum_{\lambda} (S_{0\lambda})^{2}.
\label{ZAroof}
\ee
If $X$ is non-compact, the equivariant localization is one natural way to make sense of this integral. In what follows, we present more evidence for it by showing that other methods lead to the same result.
The formula \eqref{ZAroof} has a slight generalization to the invariant of Lens spaces,
\be
Z_{\text{RW} [X]} \left( L(k,1) \right)
\; = \; \int_X e^{k \omega} \wedge \widehat{A} (TX)
\; = \; \sum_{\lambda} (S_{0\lambda})^{2} \, T_{\lambda \lambda}^k
\label{RWLens}
\ee
where $\omega$ is the symplectic form, with respect to which $U(1)_t$ is Hamiltonian. We denote by $\mu$ the corresponding moment map.
When fixed points are isolated,\footnote{We also assume that the Hamiltonian for the $U(1)_t$-action is proper, which guarantees that weights in $\CH (S^2)$ is nonnegative and weight spaces are finite dimensional. Below we consider several examples where this assumption fails.}
we have
\begin{eqnarray}
T_{\lambda \lambda} & = & t^{\mu (\lambda)} \label{STisolated} \\
(S_{0 \lambda})^2 & = & \frac{K_X^{1/2}}{\text{K-theory Euler class} (T_\lambda X)} \nonumber
\end{eqnarray}
that we illustrate in many examples below.
Here $K_X^{1/2}$ is the square root of the canonical bundle of
$X$. Since $X$ is hyper-K\"ahler, it is trivial, but not as a
$U(1)_t$-bundle.

When extending results to more general 3-manifolds, one may keep the equivariant parameter(s) $x$ but needs to set $t \to 1$ (or some other special value, {\it cf.} \cite{Dedushenko:2018bpp}), possibly using $\zeta$-function regularization after summing over $\lambda$.
Typically, this regularization is not required in examples with $U(1)_x$ symmetry, which will be the majority of our examples here, including applications to the ADO invariants and $\hat Z$-invariants.
On the other hand, examples with $U(1)_t$ symmetry but no $U(1)_x$ symmetry are the ones where most of the connections with familiar MTCs were found so far \cite{Gukov:2015sna,Dedushenko:2018bpp}.

Note, if $X$ is the Coulomb branch of a 3d $\CN=4$ theory, then there is always a ``canonical'' choice of $U(1)_t$ as the anti-diagonal subgroup of $SU(2)_R \times SU(2)_N$ R-symmetry, see appendix~\ref{app:3dtwists} for details. Moreover, in such situations, there are additional ways to compute the genus-0 equivariant Verlinde formula \eqref{ZAroof} that will be useful to us in what follows.

The simplest non-compact hyper-K\"ahler manifold is $X = \C^2$. Since it has been discussed in great detail in \cite{Gukov:2016gkn}, here we will be brief and summarize the results. In a given complex structure on $X = \C^2$, there is one tri-holomorphic and one holomorphic circle action, with a unique fixed point (the origin) and weights $(+1,-1)$ and $(+1,+1)$, respectively. Therefore, the sum over $\lambda$ in \eqref{ZAroof} consists of a single terms:
\be
(S_{00})^{2} \; = \; \frac{t}{(1-tx)(1-t/x)}
\label{SC2}
\ee
Substituting this into \eqref{VerlS} we obtain the invariants of $M_3 = S^1 \times \Sigma_g$. On more general 3-manifolds, the invariant $Z_{\text{RW} [\C^2]} (M_3)$ is equal to the Turaev torsion of $M_3$, refined by $x$. As mentioned in the Introduction, this invariant is easily computable, and will be useful to us in what follows.

\subsection*{Example: $X=T^* \cp^1$}

As our second simplest non-compact hyper-K\"ahler manifold we consider $X=T^* \cp^1$.
The Rozansky-Witten invariants for $X=T^* \cp^1$ do not seem to appear in the literature. Therefore, this will be our first non-trivial example, where we hope to say something new.

The space $X=T^* \cp^1$ can be realized as a hyper-K\"ahler quotient
\be
X \; = \; T^* \cp^1 \; = \; \mathbb{H}^2 /\!/\!/ U(1)
\label{XCP1}
\ee
{\it i.e.} as the Higgs branch of 3d $\CN=4$ SQED with $N_f=2$ charged hypermultiplets. Much like $X = \C^2$, it has two circle actions \eqref{Uones}, one tri-holomorphic and one holomorphic. The $U(1)_x \times U(1)_t$ action on $X=T^* \cp^1$ has two fixed points, $p_1$ and $p_2$, with characters
\begin{eqnarray}
TX|_{p_1} & = & x + t/x \\
TX|_{p_2} & = & x^{-1} + t x. \nonumber
\end{eqnarray}
Therefore, the equivariant localization formula for the integral of $\hat A$-genus has two contributions, and the sum over $\lambda$ in \eqref{ZAroof} has two terms:
\begin{eqnarray}
(S_{00})^{2} & = & \frac{1}{(1-x)(1-t/x)} \label{SCP1} \\
(S_{01})^{2} & = & \frac{1}{(1-x^{-1})(1-tx)}. \nonumber
\end{eqnarray}
Substituting this into the general formula \eqref{VerlS}:
\be
Z_{\text{RW} [X]} (S^1 \times \Sigma_g)
\; = \; \sum_{\lambda} (S_{0\lambda})^{2-2g},
\label{eqVerlinde}
\ee
we obtain the Rozansky-Witten invariants for $X=T^* \cp^1$ and $M_3 = S^1 \times \Sigma_g$. For example, in the genus-0 case we get
\be
Z_{\text{RW} [T^* \cp^1]} (S^1 \times S^2)
\; = \; \frac{1+t}{(1-tx)(1-t/x)}.
\label{CP1onSS2}
\ee
If we restore the overall power of $t$ that comes from $K_X^{1/2}$, we get
\be
Z_{\text{RW} [T^* \cp^1]} (S^1 \times S^2)
\; = \; \frac{t^{1/2}(1+t)}{(1-tx)(1-t/x)}.
\ee
Below we present several alternative derivations of these results.

Before we proceed, though, let us point out that the expressions obtained here behave well in the limit $t \to 1$, which we need to take when working on more general 3-manifolds. In fact, the two contributions \eqref{SCP1} from two fixed points on $X=T^* \cp^1$ become equal in the limit $t=1$, both equal to the contribution \eqref{SC2} of a single fixed point on $X = \C^2$. This suggests that, on a general 3-manifold $M_3$, the invariant $Z_{\text{RW} [T^* \cp^1]} (M_3)$ is simply the Turaev-Milnor torsion multiplied by 2:
\be
Z_{\text{RW} [T^* \cp^1]} (M_3) \; = \; 2 Z_{\text{RW} [\C^2]} (M_3)
\ee

Now, let us return to the results of the equivariant localization \eqref{SCP1}--\eqref{CP1onSS2}, refined by $t$, and reproduce them by other methods.
First, we can reproduce them by using a quantum field theory analogue of the equivariant localization, namely using the supersymmetric localization.
In order to do this, we need to realize $X=T^* \cp^1$ (or its ALF version) as a Coulomb branch in some 3d $\CN=4$ gauge theory.
Since we already know how to realize $X$ as a Higgs branch, {\it cf.} \eqref{XCP1}, we can find the desired theory by applying 3d mirror symmetry \cite{Intriligator:1996ex}. However, 3d $\CN=4$ SQED with $N_f=2$ charged hypermultiplets is self-mirror.
Therefore, $X=T^* \cp^1$ can be also realized as the Coulomb branch of the same theory,\footnote{More generally, when $X$ is the cotangent bundle of the full flag variety of type $A$, it arises as the Higgs and Coulomb branch of a self-mirror 3d $\CN=4$ theory; see {\it e.g.} \cite{Rimanyi:2019ubu} for recent discussion.}
namely $U(1)$ gauge theory with two charged hypermultiplets.

If we introduce fugacity $z$ for the $U(1)$ gauge symmetry and denote by $\tilde \CW (z)$ the twisted superpotential in 3d $\CN=4$ theory with the Coulomb branch $X$, then the sum over $\lambda$ in \eqref{eqVerlinde} can be interpreted as a sum over Bethe vacua of the gauge theory, {\it i.e.} over the critical points of $\tilde \CW (z)$.
For the theory we are interested in, the result looks like ({\it cf.} \cite{Dedushenko:2017tdw,Benini:2015noa,Closset:2016arn}):
\be
Z_{\text{RW}[X]} (S^1 \times \Sigma_g)
\; = \; \sum_{\lambda} (S_{0\lambda})^{2-2g}
\; = \; \sum_{\text{Bethe vacua}} \, \left( e^{-2 \mathcal{U}} \tilde \CW'' \right)^{g-1}
\label{ZRWSSg}
\ee
where 
$e^{-2 \mathcal{U}} = 1-t$ is the effective dilaton \cite{Nekrasov:2014xaa}, and the sum on the right-hand side is over Bethe vacua, {\it i.e.} solutions to
\be
1 \; = \; \exp \left( \frac{\partial \tilde \CW}{\partial \log z} \right) \; = \;
x \left( \frac{1 - z t^{1/2}}{t^{1/2} - z} \right)^2.
\label{BAESQED}
\ee
This equation for $z$ has two solutions, such that the two corresponding terms in \eqref{ZRWSSg} indeed reproduce the previous result of the equivariant localization with \eqref{SCP1}.
In what follows, we illustrate in a few more examples the correspondence \eqref{XMTCBethe} between the ``geometric approach'' based on the equivariant localization and the ``physical approach'' based on the supersymmetric localization, although the former tends to be easier to carry out in practice.

Note, the right-hand side of \eqref{ZRWSSg} has the standard form of the partition function in Landau-Ginzburg A-model with potential $\tilde \CW (z)$ on $\Sigma_g$ \cite{Vafa:1990mu,Melnikov:2005tk}.
The only interesting aspect is that $\tilde \CW (z)$ here is a multivalued function and, correspondingly, the equation \eqref{BAESQED} for critical points is written in the exponentiated form.
The multivaluedness of $\tilde \CW (z)$, common to all 3d $\CN=2$ theories on a circle, plays an important role in Bethe/Gauge correspondence \cite{Nekrasov:2009uh} and in 3d-3d correspondence \cite{Gadde:2013wq} where it is associated with the trigonometric nature of the corresponding integrable systems and with multivaluedness of the Chern-Simons functional, respectively.

\subsection*{Hilbert space and Hilbert series}

Another way to obtain the result \eqref{CP1onSS2} is to note that $X=T^* \cp^1$ is a resolution of the Kleinian singularity $\C^2 / \Z_2$. Since the Rozansky-Witten invariants should be independent on resolution (K\"ahler) parameters, we expect
\be
Z_{\text{RW}[T^* \cp^1]} (S^2 \times S^1)
\; = \;
Z_{\text{RW}[\C^2 / \Z_2]} (S^2 \times S^1)
\label{TCP1orbifold}
\ee
The right-hand side can be computed using the fact that, on a K\"ahler manifold $X$, the index of the Spin$^c$ Dirac operator \eqref{ZAroof} given by the integral of $\hat A$-genus counts holomorphic functions on $X$.
In the case of $X = \C^2 / \Gamma$, we have $\C [\C^2/\Gamma] = \C [\C^2]^{\Gamma}$ which, in fact, holds true for any discrete subgroup $\Gamma \subset SU(2)$ of ADE type. We shall return to this generalization shortly, after going over the details for $\Gamma = \Z_2$.

The space of holomorphic functions on $\C^2$ is, of course, infinite-dimensional. However, in evaluating the trace over this space we can use the equivariant version of the index, equivariant with respect to the symmetry $U(1)_{q_1} \times U(1)_{q_2}$ acting on $\C \times \C = \C^2$ in an obvious way,\footnote{Each $U(1)$ factor acts on the corresponding copy of $\C$ with weight $+1$.}
\be
\Tr_{\C[\C^2]} \, q_1^m q_2^n \; = \; \frac{1}{(1-q_1)(1 - q_2)}
\ee
If we parametrize $\C^2$ with complex coordinates $(z_1,z_2)$, then a holomorphic function $z_1^m z_2^n$ contributes to this index $q_1^m q_2^n$. Keeping only those generators of this ring that are invariant under $\Z_2 : (z_1,z_2) \mapsto (-z_1,-z_2)$, we obtain
\be
\frac{1 + q_1 q_2}{(1-q_1^2)(1 - q_2^2)}
\ee
Upon the change of variables
\begin{eqnarray}
t & = & q_1 q_2 \\
x & = & \frac{q_1}{q_2}  \nonumber
\end{eqnarray}
that relates the symmetry $U(1)_{q_1} \times U(1)_{q_2}$ and the corresponding equivariant parameters, $q_1$ and $q_2$, to the ones used earlier, we recover the result of the equivariant localization \eqref{CP1onSS2}.
This method, that one might call an ``orbifold approach,'' is especially effective when $X$ is (a resolution of) an orbifold singularity $\C^{2N} / \Gamma$.

More importantly, this simple example illustrates another useful interpretation of the index \eqref{ZAroof}:
\be
Z_{\text{RW}[X]} (S^2 \times S^1) \; = \;
\text{Hilbert series of } X
\label{ZRWHilbseries}
\ee
which we can use for any non-compact hyper-K\"ahler target $X$. 
Using the results of the extensive work on the Hilbert series of Coulomb branches in 3d $\CN=4$ gauge theories, one can obtain many examples of the Rozansky-Witten invariants $Z_{\text{RW}[X]} (S^2 \times S^1)$.

In particular, the mathematical definition of Coulomb branches proposed in \cite{Nakajima:2015txa} makes use of the algebra structure on the Hilbert space on $S^2$ (given by the operator product of local operators) and defines $X$ is the Spec of this commutative algebra, {\it i.e.}
\be
\CH (S^2) \; = \; \C [X].
\ee
This definition, therefore, is not only consistent with \eqref{ZRWHilbseries}, but in many cases offers various ways to compute it, \emph{e.g.} by the equivariant techniques similar to the ones used in the equivariant Verlinde formula \eqref{VerlS} (or its genus-0 version \eqref{ZAroof}, to be more precise), an explicit presentation of $X$ as a hypersurface like the one used below, {\it etc.}

\begin{table}
	\begin{centering}
		\begin{tabular}{|c||c|c|}
			\hline
			~$\phantom{\int^{\int^\int}} X \phantom{\int_{\int}}$~ & ~$M_3 = S^1 \times S^2$~ & ~$M_3 = T^3$~ \tabularnewline
			\hline
			\hline
			$\phantom{\int^{\int^\int}} \C^2 \phantom{\int_{\int}}$ & $\frac{t}{(1-tx)(1-t/x)}$ & $1$
			\tabularnewline
			\hline
			$\phantom{\int^{\int^\int}} T^* \cp^1 \phantom{\int_{\int}}$ & $\frac{t^{1/2}(1+t)}{(1-tx)(1-t/x)}$ & $2$
			\tabularnewline
			\hline
			$\phantom{\int^{\int^\int}} A_{n-1}~ \text{ALE} \phantom{\int_{\int}}$ & $\frac{t^{1/2}(1+t + \ldots + t^{n-1})}{(1 - t^{n/2} x) (1 - t^{n/2} x^{-1})}$ & $n$
			\tabularnewline
			\hline
			$\phantom{\int^{\int^\int}} D_{n}~ \text{ALE}~(n>2) \phantom{\int_{\int}}$ & $\frac{t^{1/2}(1+t^{n-1})}{(1-t^2)(1-t^{n-2})}$ & $n+1$
			\tabularnewline
			\hline
			$\phantom{\int^{\int^\int}} E_6~~ \text{ALE} \phantom{\int_{\int}}$ & $\frac{t^{1/2}(1-t^2+t^4)}{1-t^2-t^3+t^{5}}$ & $7$
			\tabularnewline
			\hline
			$\phantom{\int^{\int^\int}} E_7~~ \text{ALE} \phantom{\int_{\int}}$ & $\frac{t^{1/2}(1-t^3+t^{6})}{1-t^3-t^4+t^{7}}$ & $8$
			\tabularnewline
			\hline
			$\phantom{\int^{\int^\int}} E_8~~ \text{ALE} \phantom{\int_{\int}}$ & $\frac{t^{1/2}(1+t - t^3 - t^4 - t^{5} + t^{7} + t^{8})}{1+t-t^3-t^4-t^{5}-t^{6}+t^{8}+t^{9}}$ & $9$
			\tabularnewline
			\hline
		\end{tabular}
		\par\end{centering}
	\caption{\label{tab:examples} Invariants $Z_{\text{RW}[X]} (M_3)$ for simple $X$ and $M_3$.}
\end{table}

In Table~\ref{tab:examples} we list some examples based on \cite{Benvenuti:2006qr,Cremonesi:2013lqa,Bullimore:2015lsa}.
The ALE spaces of type $A_{n-1}$ (a.k.a. cyclic geometries) admit both a holomorphic symmetry $U(1)_t$ and a tri-holomorphic symmetry $U(1)_x$, just like $X = \C^2$ or $X=T^* \cp^1$.
Therefore, on a general 3-manifold $M_3$ one needs to take the limit $t \to 1$ and the resulting Rozansky-Witten invariant $Z_{\text{RW}[X]} (M_3)$ is a function of $x$ that encodes dependence on Spin$^c$ structures:
\be
Z_{\text{RW} [X]} (M_3;x) \; = \; \sum_{b \in \text{Spin}^c (M_3)} x^b Z_{\text{RW} [X]} (M_3;b)
\label{xvsh}
\ee
On the other hand, the geometries of type $D$ or $E$ admit only $U(1)_t$ isometry, which is {\it not} tri-holomorphic. Therefore, for these non-compact targets $X$, the invariants $Z_{\text{RW}[X]} (M_3)$ on general 3-manifolds are simply numbers, independent of $x$. Note, Table~\ref{tab:examples} also illustrates that taking the limit $t \to 1$ in such cases often requires regularization.

\subsection*{Example: the Atiyah-Hitchin space}

Not included in Table~\ref{tab:examples} is the Atiyah-Hitchin space $X = \mathbb{AH} = D_0$. The reason is that the formula for general $D_n$ does not apply to $n=0$, $1$ and $2$. However, it is easy to analyze these special cases directly.
Just like other members of the $D$ family, the Atiyah-Hitchin space has only a holomorphic symmetry $U(1)_t$ but no tri-holomorphic symmetry.
The result of the equivariant localization is, in fact, the original equivariant Verlinde formula at $k=0$.

Indeed, $X = \mathbb{AH}$ is a Coulomb branch of pure 3d $\CN=4$ super-Yang-Mills with gauge group $G=SU(2)$.
Equivalently, this is a 3d $\CN=2$ gauge theory with $G=SU(2)$ and one adjoint chiral multiplet.
Its A-twisted partition function on $M_3 = S^1 \times \Sigma_g$ has the form \eqref{ZRWSSg} with the twisted superpotental and the Bethe ansatz equation given by \cite{Gukov:2015sna}
\be
1 = \exp \left( \frac{\partial \tilde \CW}{\partial \log z} \right)
= \left( \frac{z^2-t}{1-z^2t} \right)^2.
\ee
This equation has a total of four solutions $z=\{ \pm 1, \pm i \}$, two of which should be discarded and two are related by the $\Z_2$ Weyl symmetry of $G=SU(2)$. Hence, there is only one Bethe vacuum, and we quickly learn that $Z_{\text{RW}[\mathbb{AH}]} (T^3) = 1$, in agreement with \cite{Rozansky:1996bq}. In particular, the sum over Bethe vacua in \eqref{ZRWSSg} or, equivalently, the sum over $\lambda$ in \eqref{eqVerlinde} has only one term, with
\be
\label{eq:AH}
(S_{00})^2
\; =  \; Z_{\text{RW}[\mathbb{AH}]} (S^1 \times S^2)
\; = \; - \frac{t^{3/2}}{(1+t)(1-t)^2}.
\ee
Substituting this back into \eqref{eqVerlinde}, we learn that $Z_{\text{RW}[\mathbb{AH}]} (S^1 \times \Sigma_g)$ vanishes in the limit $t \to 1$ when $g \ge 2$. This is in agreement with the fact that $Z_{\text{RW}[\mathbb{AH}]} (M_3)$ is expected to compute the Casson-Walker-Lescop invariant of $M_3$ \cite{Rozansky:1996bq}, and the Casson-Walker-Lescop invariant of $M_3 = S^1 \times \Sigma_g$ vanishes when $g \ge 2$. The case $g=0$ is more delicate because \eqref{eqVerlinde} requires regularization. If we use the $\zeta$-function regularization,
\be
\lim_{t \to 1} \frac{1}{(1-t)^2} \; = \; \zeta (-1) \; = \; - \frac{1}{12}
\ee
we obtain $\lim\limits_{t \to 1} Z_{\text{RW}[\mathbb{AH}]} (S^1 \times S^2) = \frac{1}{24}$, which is half of the value of the Casson-Walker-Lescop invariant for $S^1 \times S^2$. This suggests that the regularization of the limit $t \to 1$ leads to a normalization factor $\frac{1}{2}$. It would be desirable to understand the origin of this factor better.\footnote{This behavior at $b_1 (M_3)=1$ is similar to the behavior in the Donaldson-Witten theory on a 4-manifold of the form $M_4 = S^1 \times M_3$ encountered in \cite{Marino:1998eg}. As an effective 3d topological theory on $M_3$, the latter is expected to be precisely the Rozansky-Witten theory with the target space $X = \mathbb{AH}$ \cite{Rozansky:1996bq}.} Below we present a different computation for $X = \mathbb{AH}$ that will lead to the same result \eqref{eq:AH}.

\subsection*{Example: $SU(2)$ gauge theories with fundamental matter}

Consider the 3d $\mathcal N=4$ gauge theory with gauge group $G=SU(2)$
and $n$ fundamental hypermultiplets. It is known that its Coulomb
branch is the type $D_n$ ALF space. (See {\it e.g.}
\cite[Lemma~6.9]{Braverman:2016wma}.) It is well-known that the type
$D_n$ ALF space is a hypersurface $x^2=y^2z - z^{n-1}$
for $n\ge 1$ and $x^2 = y^2 z + y$
for $n=0$. It has a $U(1)_t$ action with $\deg x=n-1$, $\deg y=n-2$,
$\deg z=2$. The canonical bundle of $X$ is trivial. But it has a
nontrivial $U(1)_t$ action with weight $1$. Therefore we multiply the
character of the coordinate ring by its square root $t^{1/2}$.

When $n>2$, the theory is good, \emph{i.e.}, degrees of generators are
all positive. The Hilbert series is
\begin{equation}\label{eq:1}
  \frac{t^{1/2}(1-t^{2(n-1)})}{(1-t^2)(1-t^{n-2})(1-t^{n-1})} =
   \frac{t^{1/2}(1+t^{n-1})}{(1-t^2)(1-t^{n-2})}.
\end{equation}
The exponents $2$, $n-2$, $n-1$ in the denominator are degrees
of generators, and $2(n-1)$ in the numerator is the degree of the
relation. Note that our $t$ here is $t^2$ in \cite{Braverman:2016wma}.

Let us recall the monopole formula \cite{Cremonesi:2013lqa} in order
to produce this answer in a different way. For a 3d $\mathcal N = 4$
gauge theory with gauge group $G$ and hypermultiplets in a
quaternionic representation $\mathbf M$, it is given by
\begin{equation*}
  \sum_\lambda t^{2\Delta(\xi)} P_G(t;\xi),\qquad
  \Delta(\xi) = -\sum_{\alpha} |\langle\alpha,\xi\rangle|
  + \frac14 \sum_\mu |\langle\mu,\xi\rangle|,
\end{equation*}
where $\xi$ runs over cocharacters of $G$ modulo the Weyl group,
$\alpha$ over positive roots, and $\mu$ over weights of $\mathbf M$
counted with multiplicities. The term $P_G(t;\xi)$ is the Hilbert
series of the ring of $\operatorname{Stab}_G(\xi)$-invariant
polynomials over the Cartan subalgebra of the Lie algebra of $G$.

For our example, it is
\begin{equation*}
    \frac{t^{1/2}}{1-t^2} + \frac{t^{1/2}}{1-t}\sum_{i>0} t^{i(n-2)}.
\end{equation*}
The first term corresponds to the cocharacter $\xi=0$,
while the second sum corresponds to $\xi=i > 0$. 
We multiply all terms by $t^{1/2}$ as above.

If $n\le 2$, the monopole formula is \emph{not} a well-defined formal power series in $t$, as it involves both positive and negative powers. Therefore it is dangerous to compute it formally. If we nevertheless try and proceed, for $n=0$ we get
\begin{equation}\label{eq:2}
    \frac{t^{1/2}(1+t^{-1})}{(1-t^2)(1-t^{-2})}
    = - \frac{t^{3/2}}{(1-t^2)(1-t)}.
\end{equation}
This is equal to \eqref{eq:AH}.
For $n=1$, we get
\begin{equation}\label{eq:3}
    \frac{-2t^{3/2}}{(1-t^2)(1-t)}.
\end{equation}
Note that, up to an extra factor of 2, this answer is equal to that of
$D_0 = \mathbb{AH}$. In particular, the $t$-dependence is
identical. This agrees with the fact that $D_1$ is a double cover of
$D_0$ (see \emph{e.g.} \cite{Seiberg:1996nz}).

Note that the equivariant localization also gives \eqref{eq:2}--\eqref{eq:3} formally. For $n=0$, we have a single fixed point
$(x,y,z) = (0,0,0)$ at which the tangent space has weights $-1$, $2$.
For $n=1$, we have two fixed points $(x,y,z) = (\pm \sqrt{-1},0,0)$
with weights $-1$, $2$.

For $n=2$, we do not see how to treat $1-t^{n-2}$ in the
denominator. Type $D_2$ surface is $x^2 = y^2z - z$ with $\deg x=2$,
$\deg y = 0$, $\deg z=4$. Thus $\frac1{1-t^0}$ is the contribution of
$1 + y + y^2 + \dots$ to the Hilbert series. The variable $y$
corresponds to the monopole operator for the cocharacter $\xi=1$,
which is clear from the above monopole formula.
A fixed point has $(x,z) = (0,0)$, but $y$ is arbitrary. Hence the
fixed point set is $\C$, non-isolated and noncompact.

Let us return back to the case $n=4$ and reproduce \eqref{eq:1} by the
equivariant localization. The $U(1)_t$ action on $X = D_4$ has three
isolated fixed points $p_1$, $p_2$, $p_3$ and one-dimensional fixed
component $\cp^1$. Tangent spaces at $p_1$, $p_2$, $p_3$ have
characters
\begin{equation*}
  TX|_{p_i} = t^{-1} + t^2.
\end{equation*}
The conormal bundle to $\cp^1$ is $\mathcal O_{\cp^1}(2)$, twisted by
$t$. Therefore we have
\begin{equation*}
  \frac{3t^{1/2}}{(1-t^{-1})(1-t^2)}
  + t^{1/2}\chi(\cp^1,(1-t{\mathcal O_{\cp^1}(2)})^{-1}).
\end{equation*}
Expanding $(1-t{\mathcal O_{\cp^1}(2)})^{-1}$ as
$1 + t {\mathcal O_{\cp^1}(2)} + t^2 {\mathcal O_{\cp^1}(4)} +
\cdots$ and using $\chi(\cp^1,\mathcal O_{\cp^1}(k)) = k+1$, we see
that the sum is equal to \eqref{eq:1} with $n=4$.

For genus $g=1$, we integrate the altenating sum of exterior powers of
the cotangent bundle $T^*X$. The equivariant localization gives
\begin{equation*}
  3 + \chi(\cp^1,1 - \mathcal O_{\cp^1}(-2)) = 5,
\end{equation*}
where $\mathcal O_{\cp^1}(-2)$ appears as the tangent bundle to $\cp^1$.

\subsection*{Example: $X=T^* \cp^n$}

So far, we managed to reproduce the answer for $X=T^* \cp^1$ using the ``geometric approach,'' the ``physics approach,'' and the ``orbifold approach.'' All these methods easily generalize to the case of $X=T^* \cp^n$, which also admits a famous hyper-K\"ahler metric constructed by Calabi \cite{MR543218}.

The hyper-K\"ahler metric on $X=T^* \cp^n$ can be constructed as a hyper-K\"ahler quotient (see {\it e.g.} \cite{MR1463814} for a lucid review):
\be
X \; = \; T^* \cp^{n} \; = \; \mathbb{H}^{n+1} /\!/\!/ U(1).
\label{XCPN}
\ee
This construction is realized on the Higgs branch of 3d $\CN=4$ SQED with $N_f=n+1$ charged hypermultiplets:
\begin{equation}
\begin{tikzpicture}[every loop/.style={min distance=15mm},
roundnode/.style={circle, draw=black, very thick, minimum size=7mm},
squarednode/.style={rectangle, draw=black, very thick, minimum size=5mm},
]
\node[roundnode]     (gauge) at (0,0)   {$1$};
\node[squarednode]   (flavor3)    at (3,0)  {$N_f$};
\draw[black, thick] (gauge.east) -- (flavor3.west) node[midway,above]{$~$};
\end{tikzpicture}
\label{NcNfquiver}
\end{equation}
The hyper-K\"ahler quotient construction also makes manifest the symmetries of $X=T^* \cp^n$. There is a holomorphic action of $U(n+1)/\Z_{n+1}$, which contains a $U(1)$ subgroup that acts by ``stretching the fibers'' and $SU(n+1)/\Z_{n+1}$ that acts tri-holomorphically.
Let $\mathbb{T}$ be the maximal torus of this tri-holomorphic symmetry and $x = (x_1, \ldots, x_{n+1}) \in \mathbb{T}_{\C}$ denote the corresponding equivariant parameters. As usual, we denote by $U(1)_t$ the holomorphic circle action, which is not tri-holomorphic. Then, there are a total of $n+1$ isolated fixed points $p_i$ on $X=T^* \cp^n$, with
\be
TX|_{p_i} \; = \; \sum_{j \neq i} x_i/x_j + t x_j / x_i.
\label{TCPNweights}
\ee
And, by the equivariant localization formula, the integral of the $\hat{A}$-genus has the form \eqref{ZAroof} with
\be
(S_{0 \lambda})^2
\; =  \;
\frac{1}{\prod_{j \neq i} (1-x_i/x_j)(1-t x_j/x_i)} \,,
\qquad \lambda \equiv i-1 = 0, \ldots, n.
\label{SCPN}
\ee
Substituting this back into \eqref{eqVerlinde} we obtain the general answer for $M_3 = S^1 \times \Sigma_g$. Note, in the case $n=1$ this result agrees with the previously obtained formula for $X=T^* \cp^1$.
Also note that, in the limit $t \to 1$ that needs to be taken on a general 3-manifold, each of the $n+1$ ``matrix elements'' in \eqref{SCPN} is a product of $n$ copies of the corresponding expressions for $X = \C^2$, with $x = x_i/x_j$, $i$ fixed and $j \ne i$.

It is straightforward, though more laborious, to reproduce these results via the physics approach \eqref{ZRWSSg}--\eqref{BAESQED} based on supersymmetric localization in 3d $\CN=4$ gauge theory with the Coulomb branch $X=T^* \cp^n$.
Such theory, related to \eqref{NcNfquiver} by 3d mirror symmetry, is a $U(1)^{n}$ gauge theory with matter content represented by a linear quiver \cite{Intriligator:1996ex}. For example, for $n=2$ ({\it i.e.} $N_f=3$) we have
\begin{equation}
\begin{tikzpicture}[every loop/.style={min distance=15mm},
roundnode/.style={circle, draw=black, very thick, minimum size=7mm},
squarednode/.style={rectangle, draw=black, very thick, minimum size=5mm},
]
\node[squarednode]     (flavor1) at (0,0)   {$1$};
\node[roundnode]     (gauge1) at (3,0)   {$1$};
\node[roundnode]     (gauge2) at (6,0)   {$1$};
\node[squarednode]   (flavor2) at (9,0)  {$1$};
\draw[black, thick] (flavor1.east) -- (gauge1.west) node[midway,above]{$~$};
\draw[black, thick] (gauge1.east) -- (gauge2.west) node[midway,above]{$~$};
\draw[black, thick] (gauge2.east) -- (flavor2.west) node[midway,above]{$~$};
\end{tikzpicture}
\end{equation}

We also note that the non-compact space $X=T^* \cp^n$ is a resolution
of the minimal nilpotent cone, which is also the reduced moduli space
of $1$-instantons of $SU(n)$ on $\R^4$. One can reproduce
\eqref{SCPN}, or rather its sum over $\lambda$ from the monopole formula.
See Section~3.2 of \cite{Cremonesi:2013lqa}.

\subsection*{Relation to non-semisimple MTCs}

Now, after working out a handful of explicit examples, we are ready to start making first connections to non-semisimple\footnote{Inspired by \cite{MR3708086}, we are tempted to call non-semisimple modular categories ``logarithmic MTCs'' (or log-MTCs for short) in order to highlight connections with logarithmic vertex algebras, which we barely touch in this work but expect to play a major role in the future developments.} MTCs and related ``TQFTs'' that don't obey Atiyah's axioms. One lesson from connecting such theories to Rozansky-Witten invariants with non-compact $X$ is that they secretly are based on infinite-dimensional spaces of states (even though it may not be obvious from the explicit construction based on a non-semisimple MTC with finitely many projective modules).

Non-semisimple modular categories naturally come from logarithmic vertex algebras, as their representation categories. For example, one prominent family of log-VOAs, associated with $\frak{g} = \frak{sl}_2$ Lie algebra\footnote{Higher-rank analogues also exist, although they are less studied. They should be related to our discussion in Section~\ref{sec:Zhat}.} and labeled by $p \in \Z_+,$ is the so-called $(1,p)$ triplet model.
It has the central charge $c = 13 - 6p - 6p^{-1}$ and, via its tensor category, relates to the restricted quantum group $\bar \CU_q (\frak{sl}_2)$ at the even primitive root of unity $q = e^{\frac{2\pi i }{2p}} = e^{\pi i /p}$.
The first non-trivial member of this family, with $p=2$, overlaps with another class of log-VOAs and log-MTCs, the family of so-called symplectic fermions with $c=-2N$.
At this value of $p$, it leads to topological invariants of knots and 3-manifolds \cite{MR1164114,MR2466562,MR3286896}, such that $Z (S^1 \times S^2) = \frac{1}{(1-x)(1-1/x)}$, in which we can quickly recognize the Rozansky-Witten invariant for $X = \C^2$, {\it cf.} \eqref{SC2}.
Although the corresponding non-semisimple MTC has $3p-1=5$ objects, there is only one fixed point on $X = \C^2$. We interpret this as a suggestion for the following:
\begin{conj}\label{conj:ss}
Even when $Z_{\text{RW}[X]} (M_3)$ computes invariants associated with a non-semisimple MTC, that we denote $\text{MTC} [X]$, the sum over $\lambda$ in the equivariant localization formula \eqref{eqVerlinde},
\be
Z_{\text{RW} [X]} (S^1 \times \Sigma_g)
\; = \; \sum_{\lambda} (S_{0\lambda})^{2-2g}
\ee
or, equivalently, the sum over Bethe vacua in \eqref{ZRWSSg} runs over \emph{simple} modules in the semisimplification of $\text{MTC} [X]$.
In other words,
\end{conj}
\be
\{ \lambda \} \qquad \xleftrightarrow[~~~~~]{~~~~~} \qquad \text{simple modules of MTC}[X]^{ss}.
\ee
This identification is also natural from the physics perspective where, even in a larger class of 3d $\CN=2$ theories, there is a relation between twisted partition functions and the category of line operators \cite{Gukov:2016gkn}. The state-operator correspondence relates line operators in a 3-dimensional theory to states on $T^2$.
While both sets are huge in a typical theory of interest (with continuous spectrum in flat space), in a 3d theory on a circle there is a particular set of states associated with the critical points of $\tilde \CW$.
We expect the corresponding line operators to be the simple objects of $\text{MTC} [X]^{ss}$,
\be
QK_{U(1)_t} (X) \; \cong \; K^0 \left( \text{MTC} [X]^{ss} \right).
\ee
Here, $QK_{U(1)_t} (X)$ is the quantum equivariant K-theory of $X$, illustrated in appendix~\ref{app:ktheory} for $X = T^* \cp^1$.

It would be also of interest to study the relation to modularity and boundary chiral algebras in 3d theories \cite{Cheng:2018vpl,Costello:2018fnz,Costello:2020ndc}. We leave this to future work.

\section{A general proposal}
\label{sec:general}

In the previous section we discussed several methods that allow to compute the analogue of the Verlinde formula in the Rozansky-Witten theory with a non-compact target $X$, effectively reducing it to a sum over a finite set of ``states'' $\{ \lambda \}$.
The latter effectively ``package'' the original infinite-dimensional state space \eqref{HRW} into a finite-dimensional one.
Here we take this interpretation one step further and, following \cite{Dedushenko:2018bpp}, propose to use this structure to evaluate $Z_{\text{RW}[X]} (M_3)$ on more general 3-manifolds, such as plumbed manifolds or surgeries on knots.

\begin{figure}[ht]
	\centering
	\includegraphics[scale=2.5]{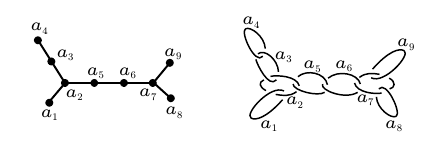}
	\caption{The surgery presentation (Kirby diagram) of a plumbed 3-manifold can be conveniently represented by a decorated graph.}
	\label{fig:plumbing-example}
\end{figure}

Let us start with plumbed 3-manifolds, that can be conveniently labeled by graphs (possibly, with loops). And, to avoid clutter, let us assume that $X$ has one tri-holomorphic symmetry $U(1)_x$. Generalization to larger tri-holomorphic symmetry is straightforward and, if $X$ happens to admit a larger tri-holomorphic symmetry, then one can pick any $U(1)$ subgroup for the purposes of the present discussion.
In this setup, our first small Lemma is that on a plumbed manifold the partition function of a Rozansky-Witten theory with $| \{ \lambda \} | = 1$, {\it i.e.} with $Z_{\text{RW}[X]} (T^3)=1$, is given by
\be
Z_{\text{RW}[X]} (M_3) \; = \;
\sum_{m_i \in \Z} \int \prod_{i \in \text{vertices}} \frac{dx_i}{2\pi i x_i} x_i^{\sum_j Q^{ij} m_j} g(x_i)^{\deg (i) - 2} x_i^{b_i}
\label{ZRWplumbed}
\ee
where $x_i \in \C^*$ is associated to the $i$-th vertex of the plumbing graph, $Q$ is the adjacency matrix, and
\be
\frac{1}{g^2(x)}
\; = \; (S_{00})^{2}
\; = \; Z_{\text{RW} [X]} \left( S^1 \times S^2 \right).
\ee
Indeed, as shown in \cite{Gukov:2016gkn}, the integral \eqref{ZRWplumbed} is invariant under the Kirby moves for {\it any} choice of the function $g(x)$ and the result depends on the background for $U(1)_x$ symmetry or, equivalently, on $b \in \text{Spin}^c (M_3)$, {\it cf.} \eqref{xvsh}.
The choice of $g(x)$ is determined by the choice of $X$.
In other words, in the present setup the dictionary \eqref{XTQFT} reads
\be
X \qquad \xleftrightarrow[~~~~]{~~~~} \qquad g(x)
\ee
For example, $g^2(x) = -x^{-1}(1-x)^2$ leads to the Turaev-Milnor torsion $\Delta (M_3)$, whereas $g^2(x) = -x (1-x)^{-2}$ leads to the inverse torsion $\frac{1}{\Delta (M_3)}$ relevant to \cite{Gukov:2019mnk}.
A large class of examples that produce \eqref{ZRWplumbed} with different $g(x)$ comes from $X$ with one isolated fixed point $p \in X$, such that $TX|_p \cong \C^{2n}$ has weights $(w_1,-w_1,w_2,-w_2, \ldots)$ under $U(1)_x$ action.\footnote{The weights must come in pairs because the tri-holomorphic symmetry $U(1)_x$ must preserve the holomorphic symplectic 2-form which, in local coordinates $(z_i,z_i')$ on $TX|_p \cong \C^{2n}$, looks like $\Omega = \sum_i dz_i \wedge dz_i'$.}
In this class of examples --- which includes the Lee-Weinberg-Yi metrics, the Taubian-Calabi metrics, and their generalizations \cite{MR1463814} --- we have
\be
g^2 (x) \; = \; \prod_{i=1}^{\frac{1}{2} \dim_{\C} X} (1-x^{w_i})(1-x^{-w_i}).
\ee
Note, this applies even to the case $\dim_{\C} X = \infty$, something that will be handy to us in Section~\ref{sec:Zhat}.

More generally, if $X$ has multiple isolated fixed points $p_s$ with the corresponding weights $\{ w_i (p_s)\}$, like a Nakajima quiver variety, we have
\be
Z_{\text{RW}[X]} (M_3) \; = \;
\sum_{s} \prod_{i=1}^{\frac{1}{2} \dim_{\C} X} \Delta (M_3, x^{w_i (p_s)})
\label{ZRWAlex}
\ee
which, for $b_1 (M_3) = 1$, reduces to the result of \cite{Blau:2000iy,Habegger:1999yp}.
For plumbed manifolds, it also reduces to the previous expression \eqref{ZRWplumbed} or, equivalently, a formula in terms of $S$ and $T$ matrix elements (see {\it e.g.} \cite{Dedushenko:2018bpp}).
The structure on the right-hand side of \eqref{ZRWAlex} also appears in the literature on generalized Alexander invariants and can be viewed as a TQFT reason why such generalizations often end up related to the Alexander polynomial \cite{MR1822139,MR2153122,MR3463851,MR3857661}.
(See, however, the discussion in Section \ref{sec:Zhat}).

\subsection*{Nakajima quiver varieties}

Three-dimensional mirror symmetry \cite{Intriligator:1996ex} allows to describe the Coulomb branch of a 3d $\CN=4$ theory as a quiver variety.
Given a quiver $\mathcal{Q}$ let
\be
X \; = \; (T^*\text{Rep} \, \mathcal{Q}) /\!/\!/ G_{\mathcal{Q}}
\ee
be the corresponding Nakalima quiver variety. The variety $X$ is equipped with a set of tautological equivariant bundles $\{V_a\}$, $\{ W_a \}$ with the action of $\prod_a GL(V_a) \times \prod_a GL(W_a) \times \mathbb{C}^*_{t}$ which correspond to gauge, flavor, and the $U(1)_t$ symmetry, respectively.
For convenience, we denote the equivariant characters of these bundles by the same letters, $V_a$ and $W_a$.
Then, in terms of gauge and flavor equivariant parameters,
\be
W_a \; = \; \sum_{i \in N_f^{(i)}} x_{a,i}
\ee
\be
V_a \; = \; \sum_{i \in N^{(i)}} z_{a,i}
\ee
and the character of a tangent bundle can be written as
\be
TX = M + t^{-1} M^* - (1+t^{-1}) \sum_a V_a V^*_a
\ee 
where
\be
M = \sum_{a \to b} V_a V_b^* + \sum_a V_a W_a^*.
\ee
For example, for the two-node quiver that corresponds to $X=T^* \cp^n$ we reproduce \eqref{TCPNweights}.
Note, the proposal here implies that, on a general 3-manifold, $Z_{\text{RW}[T^* \cp^n]} (M_3)$ is simply a multiple of the Turaev-Milnor torsion when $n=1$, but not when $n>1$.

\section{Affine Grassmannians, \texorpdfstring{$\hat Z$}{Zhat}, and logarithmic knot invariants}
\label{sec:Zhat}

In this section, we consider $q$-series invariants of 3-manifolds, $\hat Z (M_3)$, and relate them to the Rozansky-Witten invariants, on the one hand, and to the Akutsu-Deguchi-Ohtsuki (ADO) knot invariants, on the other.
In particular, this will require working with target spaces $X$ which are not only non-compact, but also infinite-dimensional.

For $G=SU(2)$, the $q$-series invariant $\hat Z (M_3)$ provides a $q$-deformation of the inverse Turaev-Milnor torsion~\cite{Chun:2019mal}, in the sense that $\hat Z (M_3) \vert_{q \to 1} = \frac{1}{\Delta (M_3)}$. As such, it has many properties similar to those of $\Delta (M_3)$, {\it e.g.} dependence on Spin$^c$ structure, aspects of cutting and gluing formulae, {\it etc.} Since every 3-manifold can be obtained via surgeries on knots and links, of particular interest are knot complements, for which
\be
F_K (x,q) \; := \; \hat Z \left( S^3 \setminus K \right)
\label{FKcomplement}
\ee
is basically a $q$-deformation of the inverse Alexander polynomial \cite{Gukov:2019mnk}. Here, the Alexander variable $x$ encodes the dependence on Spin$^c$ structure, {\it cf.} \eqref{xvsh}. For example,
\be
F_{{\bf 3_1}} (x,q) \; = \;
\frac{q}{2} \sum_{m=1}^{\infty} \epsilon_m (x^{\frac{m}{2}} - x^{- \frac{m}{2}}) \, q^{\frac{m^2-1}{24}}
\label{FKtrefoil}
\ee
where
\be
\label{eq:epstrefoil}
\epsilon_m = \begin{cases}
	-1 & \operatorname{if } \ m \equiv 1 \ \operatorname{ or } \ 11 \!\! \pmod{12}\\
	+1 & \operatorname{if } \ m \equiv 5 \ \operatorname{ or }\  7 \!\! \pmod{12}\\
	0 & \operatorname{otherwise.}
\end{cases}
\ee
Note, that $\hat Z$-invariants, here for $G=SU(2)$, depend on a choice of the group $G$, the variable $q$, a 3-manifold $M_3$ (possibly, with boundary), and on Spin$^c$ structure (or abelian flat connection) encoded in $b$ or $x$-dependence, related via \eqref{xvsh}. Since the full notation $\hat Z_b (M_3, G,q)$ or $F_K^G (x,q)$ can look a little bulky, we often suppress dependence on some of the data when it is reasonably clear from the context.
The higher-rank versions of $\hat Z (M_3)$ and $F_K$ are discussed in \cite{Park:2019xey} and will be useful to us below.

We wish to re-formulate the $\hat Z$-invariants in the framework of \cite{Dedushenko:2018bpp}, where topological twists of 4d $\CN=2$ Argyres-Douglas theories on 3-manifolds were studied from the perspective of the Rozansky-Witten theory with the target space $X = \CM_H (G,C)$, the moduli space of $G$-Higgs bundles on a curve $C$ with wild ramification~\cite{Fredrickson:2017yka}.\footnote{This relation between twisted Argyres-Douglas theories and Rozansky-Witten invariants is conceptually the same as (and builds on) the original relation between Donaldson-Witten theory on 3-manifolds and the Rozansky-Witten invariants \cite{Rozansky:1996bq} (see also \cite{Marino:1998eg,Blau:2000iy}).}
These invariants of 3-manifolds can be understood as twisted partition functions of 6d $(0,2)$ theory on
\be
M_3 \times S^1 \times C
\label{fivebrane}
\ee
which has the same form as the physical setup for the $\hat Z$-invariants \cite{Gukov:2016gkn,Gukov:2017kmk}, except the latter requires a particular and somewhat peculiar choice of $C = D^2$. In this special case, $C$ enjoys the action of an extra rotation symmetry, that we denote $U(1)_q$ since it gives rise to $q$-dependence of $\hat Z$-invariants.\footnote{Another special choice of $C$ with the same property is $C = \cp^1$; in this case, the corresponding 3-manifold invariant also depends on $q$ and computes the index of 3d $\CN=2$ theory $T[M_3]$. Examples of such computations can be found in~\cite{Gukov:2017kmk,Chung:2019khu} and will be helpful to us below.}
This leads to a natural proposal that $\hat Z$-invariants of $M_3$ can be equivalently defined as Rozansky-Witten invariants with non-compact and infinite-dimensional target $X = \CM_H (G,D^2)$. The main goal, then, is to understand the geometry of $X = \CM_H (G,D^2)$ and what model of this space leads to the Rozansky-Witten invariants that match previous computations of $\hat Z (M_3)$.

When $C$ is compact, reducing the 6d $(0,2)$ theory on $S^1 \times C$ leads to a 3d $\CN=4$ theory $T[S^1 \times C]$ given by a sigma model with target $\CM_H (G,C)$. There are different mathematical models for this space. For example, one can identify it with the Hitchin moduli space, the moduli space of solutions to a system of PDEs known as the Hitchin's equations \cite{MR887284}.
Another good model is given by the cotangent stack $T^* \mathrm{Bun}_{G_\C}(C)$ to the moduli stack of holomorphic $G_{\C}$-bundles. 
In many cases, adopting either models yields identical results, but when the genus of $C$ is small the latter becomes a better model \cite{Andersen:2016hoj}.
We shall see this more explicitly here as well.

For the particular choice $C = D^2$ relevant to us here, there are also different models for the space $\CM_H (G,D^2)$. Conceptually, this moduli space can be identified with $T^*\text{Gr}_{G_{\C}}$, the cotangent bundle to the affine Grassmannian $\mathrm{Gr}_{G_{\C}}$.
%
This motivates the following conjecture, for which we will provide some evidence further below:
\begin{conj}[$\hat Z$ as Rozansky-Witten invariants]\label{conj:ZRW}
	\be
	\hat Z (M_3) \; = \; Z_{\text{RW}[X]} (M_3)
	\label{ZRWeq}
	\ee
	with $X = \CM_H (G,D^2)=\text{``}T^*\text{Gr}_{G_{\C}}\text{''}$.
\end{conj}

Here, $T^*\text{Gr}_{G_{\C}}$ appears in quotation marks because, just like when $C$ is a compact Riemann surface, one needs to choose a suitable mathematical model.
There are two natural models for $T^*\text{Gr}_{G_{\C}}$ that are supposed to be equivalent. One is the algebraic model\footnote{This is slightly larger than the usual model, \emph{e.g.} $\underline{\mathfrak{u}}$ in Section 7 of \cite{MR2135527}. But the difference is not essential for many purposes.}
\begin{equation}
\CT \; := \; G_{\C}(\CK)\times_{G_{\C}(\CO)}{\frak{g}_{\C}(\CO)},
\end{equation} 
where $G_{\C}(\CK):=G_\C((z))$ and $G_{\C}(\CO)=G_{\C}[[z]]$.
The second model for $T^*\text{Gr}_{G_{\C}}$ is
\begin{equation}
LG_{\C} \; := \; \mathrm{Map}(S^1,G_{\C})/G_{\C}.
\end{equation}
This space appears by considering Hitchin's equation on the actual
disk $D^2 = \{z\in\C \mid |z|<1\}$ with Dirichlet boundary condition
\cite{MR1165874}. In particular, it is hyper-K\"ahler and
diffeomorphic to the moduli space of Higgs bundles over $D^2$.

The former model uses the formal disk $D = \operatorname{Spec}\C[[z]]$ and might be easier to work with, as it is a relatively familiar space in geometric representation theory. Many of its properties are already well known and it is closely related to the construction of Coulomb branches of 3d $\CN=4$ theories \cite{Braverman:2016pwk,Braverman:2016wma}, which recently attracted a lot of attention. In fact, a crucial ingredient in this construction of Coulomb branches is the moduli space $\CR$ of Higgs bundle on a ``raviolo,'' two disks glued along a punctured disk.
This moduli space can be identified with a closed subvariety of $\CT$, given by\footnote{For more details, see {\it e.g.}~page 6 in \cite{Nakajima:2017bdt}.}
\begin{equation}
\CR \; = \; \{(g,s)\in \CT|gs\in\frak{g}_{\C}(\CO)\}.
\end{equation}
Similarly, one can construct the Hitchin moduli space associated with any Riemann surface from $\CT$ as a quotient. From this point of view, $\CT$ is the fundamental building block for all the other Hitchin moduli spaces. 

Although $\CT$ is infinite-dimensional, it has many good properties. For example, it enjoys a $U(1)_q \times U(1)_t$ action coming from the rolation of $D^2$ and the rotation of the cotangent fiber. This action has well-understood fixed points, which, in principle, enables one to perform direct computation in many cases via localization. More importantly, it seems to capture the relevant KK-modes that contribute to the computation of $\hat Z$-invariants. This should be contrasted with finite-dimensional moduli spaces of semi-stable Higgs bundles over a compact Riemann surface, which do not always capture all physical degree of freedom that can contribute to the partition functions. Once we upgrade these moduli spaces to stacks, they become more complicated than $\CT$. For example, it is no longer clear how to talk about hyper-K\"ahler geometry and how to define the Rozansky-Witten theory.  

We now provide some evidence for the Conjecture~\ref{conj:ZRW} by looking at the simplest 3-manifold, the three-sphere.

\subsection*{$M_3 = S^3$ and generalizations}

As we discussed earlier, when $M_3 = L(k,1)$ one can turn on an additional parameter associated with the $U(1)_t$ symmetry, and the Rozansky-Witten theory partition function can be identified with the equivariant Verlinde formula~\cite{Gukov:2015sna}. The latter can be interpreted as the equivariant index of a line bundle over the Hitchin moduli space $X=\CM_H(C,G)$: 
\begin{equation}
Z_{\text{RW}[X]} (L(k,1)) \; = \; \mathrm{Index}_{S^1}(X,\CL^k;t) \,.
\end{equation}
In both models for $T^*\mathrm{Gr}_{G_{\C}}$, one can make sense of the line bundle $\CL$, and the Conjecture~\ref{conj:ZRW} becomes
\begin{equation}
\hat{Z}(L(k,1);q,t) \; = \; \mathrm{Index}_{S^1 \times S^1}(T^*\mathrm{Gr}_{G_{\C}},\CL^k;q,t) \,.
\label{Lk1RW}
\end{equation}
For $k=1$, that is for $M_3=S^3$, the left-hand side is \cite{Gukov:2017kmk}:
\begin{equation}
\hat{Z}(S^3;q,t)=\prod_{i=1}^{\mathrm{rank}\,G} \frac{1}{(q^{d_i}t^{d_i};q)_\infty}
\end{equation}
where $d_i$ denotes the degree of the $i$-th fundamental invariant of $\frak{g}$. The right-hand side of this formula is exactly what appears in Theorem 11.12 and in equation (12.4) of \cite{MR2415401}, where a closely related computation over the affine Grassmannian is performed. It would be interesting to study this more carefully and to generalize it to $k>1$.

This expression also appears as the generating function of the
Poincar\'e polynomials of the intersection cohomology groups of
$G$-instanton moduli spaces on $\R^4=\C^2$ when $G$ is of type $ADE$ (see
Th.~7.10 in \cite{BFG}, Section~6 in \cite{braverman-2007}). Its
natural generalization to $k > 1$ is to consider
$G$-instanton moduli spaces on $\C^2/(\Z/k\Z)$ \cite{braverman-2007}.
The relation between $\hat{Z}$ and instanton moduli spaces was also noticed in \cite{Gukov:2016gkn}.
If we replace $T^*\mathrm{Gr}_{G_{\C}}$ by the cotangent bundle
$T^* X$ of the usual finite-dimensional flag variety $X = G/T$, we
have a finer relation between spaces of sections of line bundles and
intersection cohomology groups of singular $G$-monopole moduli spaces
on $\R^3$ (a.k.a.\ affine Grassmannian slices): one can reconstruct
$T^* X$ as $\operatorname{Proj}$ of direct sum of intersection
cohomology groups \cite{MR2053952}. This construction can be regarded
as a precusor to the mathematical approach to Coulomb branches of
gauge theories. See \cite{2017arXiv170602112B} for more detail.
Therefore the coincidence between \eqref{Lk1RW} and the Poincar\'e
polynomial suggests a possibility of yet another model of
$T^*\mathrm{Gr}_{G_{\C}}$.

For $M_3 = L(k,1)$ with $k>1$, one needs to understand how the dependence of $\hat Z_b (M_3)$ on Spin$^c$ structure $b$ should enter the right-hand side of \eqref{Lk1RW}. It is likely that the choice of $b \in \text{Spin}^c (M_3)$ corresponds to replacing $\CL^k$ with a different sheaf $\CF_{k,b}$ that depends on $b$.
It is expected to be equivariant, so that the localization theorem still applies. One way to understand this better is by gluing two copies of $C=D^2$ in \eqref{fivebrane} to make a $\cp^1$, and then use the fact \cite{Gukov:2017kmk} that the equivariant Verlinde formula for $C = \cp^1$ can be written as a sum of products of $\hat{Z}_b (M_3)$. 
	
Another generalization that can help to test the Conjecture~\ref{conj:ZRW} is to include a line operator in $M_3$ labeled by a $G$-representation $R_\lambda$; it will also be a line in the solid torus $D^2 \times_q S^1$ wrapping its core. Then, the right-hand side of \eqref{Lk1RW} will be replaced by 
\begin{equation}
\mathrm{Index}_{S^1 \times S^1}(T^*\mathrm{Fl}_{G_{\C}},\CL_{k,\lambda};q,t)
\end{equation}
where $\mathrm{Fl}_{G_{\C}}$ is the affine flag variety and $\CL_{k,\lambda}$ is a line bundle that become the usual line bundle $L_\lambda$ after pulling back to the ordinary, finite-dimensional flag variety. From physics, we expect that such indices are closely related to the Macdonald polynomials.

\subsection*{The geometry of $q$}

The tri-holomorphic symmetry $U(1)_q$ is a genuine symmetry of the Rozansky-Witten model, and just like similar tri-holomorphic symmetries considered previously, one can refine the partition function on any three-manifold $M_3$ by turning on a flat $U(1)_q$ background field. This is how the dependence on $b \in \text{Spin}^c (M_3)$ naturally appears in~\eqref{ZRWeq}.\footnote{When $b_1(M_3)>0$, This also gives rise to $U(1)$-valued parameters, which can be complexified to be the $\C^*$-valued $x$ variables that we have seen before \textit{e.g.}~in \eqref{FKcomplement}.}
Here, we wish to comment on the $q$-dependence in \eqref{ZRWeq} by considering a similar but simpler example. This is the example of genus-0 equivariant Verlinde formula, which is known to have a refinement by $q$.

The equivariant Verinde formula can be identified with the index
\begin{equation}
\CI (C,G;k,t) \; = \; \mathrm{Index}(\CM(C,G),\CL^k \otimes S_tT_{\CM})
\end{equation}
where $\CM (C,G)$ is the moduli stack of $G$-bundles on $C$, $\CL$ is the ``determinant line bundle,'' and $S_tT_{\CM}$ is the total symmetric power of the tangent sheaf of $\CM$.
For $C=\cp^1$, one can ask whether the symmetry of $\cp^1$ can be used to introduce another parameter $q$. Physics suggests that this is indeed possible, as $\CI (C,G;k,t)$ can now be identified with a topologically twisted index of a 3d $\CN=2$ theory, which can be refined by a parameter $q$. Specifically \cite{Gukov:2017kmk},
\begin{equation}
\CI (\cp^1,SU(2);k=1,t,q) \; = \; (1-tq^{-2})(1-t)(1-tq^2) \,.
\label{IS3SU2}
\end{equation} 
On the other hand, we expect that, for $k>2h-2$, there is no dependence on $q$. This follows from a result in Section 7 of \cite{Andersen:2016hoj} that claims the contributions from the unstable strata vanish in this case. Then, only the substack of semi-stable bundles $\CM^{\text{ss}}(\cp^1)=BG$ contributes to the index of $\CL^k$, but $BG$ does not carry a non-trivial action of the $SU(2)$ symmetry of $\cp^1$. 

Unstable strata in $\CM(\cp^1)$ are labeled by co-characters of $G$. Given a co-character $\xi$, the contribution from the corresponding stratum is \cite{Andersen:2016hoj}
\begin{align}
\CI_{\xi}(t)&:=|W_\xi|^{-1}(-1)^{\#}(1-t)^{\mathrm{rk}\,G}\cdot \mathrm{Inv}\left[\prod_{\alpha>0}\frac{(1-te^{-\alpha})^{\alpha(\xi)+1}}{(1-te^{\alpha})^{\alpha(\xi)-1}}
 \cdot e^{(k+h)\langle\xi,\cdot\rangle-2\rho}\prod_{\alpha>0} (1-e^{\alpha})^2\right].
\end{align}
Here ``Inv'' is the operation of taking the zero mode piece of a function on the Cartan $T\subset G$.\footnote{The $(-1)^{\#}$ phase factor here has a precise expression, but we will suppress it to avoid clutter.}
This is very similar to expressions that show up in supersymmetric localization computations. In fact, the above expression can be directly compared with the gauge theory computation leading to \eqref{IS3SU2} which naturally incorporates $q$. Therefore one can attempt to reconstruct the $q$-refinement for the above expression. One arrives at \begin{align}\nonumber
\CI_{\xi}(q,t)&:=|W_\xi|^{-1}(-1)^{\#}(1-t)^{\mathrm{rk}\,G}q^{\rho(\xi)} \cdot \\& \mathrm{Inv}\left[e^{(k+h)\langle\xi,\cdot\rangle}\cdot\prod_{\alpha>0}\frac{(1-q^{\alpha(\xi)}e^{\alpha})(1-q^{\alpha(\xi)}e^{-\alpha})}{(tq^{1-\alpha(\xi)/2}e^{\alpha};q)_{\alpha(\xi)-1}(q^{1+\alpha(\xi)/2}te^{-\alpha};q)_{-\alpha(\xi)-1}}
  \right]\label{Iqt}
\end{align}
where we have used the following convention for $q$-Pochhammers
\begin{equation}
(z;q)_{-n}^{-1}=(zq^{-n};q)_n.
\end{equation}

Now one can try to identify the coherent sheaf whose Euler characteristics is computed by \eqref{Iqt}. This will shed light on the origin of the $q$-parameter in the Rozansky-Witten model interpretation of $\hat{Z}$. We hope to return to this in future work.

Another remark is that, while $\CI$ corresponds to the topologically twisted index in physics, and there is another quantity known as the superconformal index. The latter should also have a similar decomposition into a sum over a expression similar to \eqref{Iqt} but with finite $q$-Pochhammers  replaced with infinite ones. 

\subsection*{Relation to the ADO invariants}

As we emphasized in the above, the $q$-series invariant $\hat Z (M_3,SU(2),q)$ can (and should) be thought of as the quantum group invariants associated with $\CU_q (\frak{sl}_2)$ at a generic value of the parameter $|q|<1$.
Once the dependence on the Spin$^c$ structure is removed, {\it viz.}~summed over with particular weights, the limit $q \to e^{2\pi i/k}$ is related to the Witten-Reshetikhin-Turaev (WRT) invariants at (renormalized) level $k$.
Curiously, the weights in this sum are $S$-matrix elements of a non-semisimple MTC \cite{Gukov:2016gkn}. Moreover, from the viewpoint of the underlying representation theory, the relation to WRT invariants is a two-step process (see {\it e.g.} \cite{Feigin:2005zx,Cheng:2018vpl} for a review):

\begin{enumerate}
\item
taking the limit $q \to e^{2\pi i/k}$ leads to a non-semisimple MTC, and then

\item
a further semisimplification yields the Verlinde category, which encodes the algebraic structure of cutting and gluing relations of WRT invariants.

\end{enumerate}

\noindent
All these clues suggest that, perhaps, even a simpler and more natural relation should involve $\hat Z$-invariants at roots of unity and topological invariants associated with non-semisimple MTCs before taking the second step and before removing the dependence on Spin$^c$ structures.

Here, we present some evidence that such relation indeed holds true for knot complements and leave the study of more general 3-manifolds to future work.
The non-semisimple MTC associated with $\CU_q (\frak{sl}_2)$ at roots of unity is precisely the one described earlier, also related to logarithmic CFTs.
The corresponding knot invariants were studied by Akutsu-Deguchi-Ohtsuki \cite{MR1164114} and, in a closely related work, by Murakami-Nagatomo \cite{MR2466562}.
In particular, Akutsu-Deguchi-Ohtsuki introduced a family of polynomial knot invariants, that we denote $\text{ADO}_p (x;K)$, associated with the quantum group $\CU_q (\frak{sl}_2)$ at the even $2p$-th root of unity. These polynomial knot invariants generalize the Alexander polynomial $\Delta_K (x)$, in a sense that
\be
p=2: \qquad
\text{ADO}_2 (x;K) \; = \; \Delta_K (x)
\label{ADOAlex}
\ee
and, as expected, emerge from the $\hat Z$-invariants for knot complements \eqref{FKcomplement} in the limit $q \to \zeta_p \equiv e^{2\pi i/p}$.

\begin{conj}[ADO from $q$-series invariants]\label{conj:FKADO}
For any knot $K$,
\end{conj}
\be
\left. F_K (x,q) \right|_{q = \zeta_p} \; = \;
\frac{\text{ADO}_p (x/\zeta_p;K)}{\Delta_K (x^{p})}
\,, \qquad \text{where} \quad \zeta_p = e^{2\pi i/p} \,.
\label{FKADOconj}
\ee
Combining the recent results from \cite{SPark} and \cite{Willetts}, it is not hard to prove this conjecture for positive braid knots (as well as other examples considered in \cite{SPark}). The simplest non-trivial example of a non-positive braid knot is the figure-8 knot $K={\bf 4_1}$ that will be one of our examples below.

The Conjecture~\ref{conj:FKADO} is also motivated from the physics perspective as follows. In the setup of~\cite{Gukov:2016gkn,Gukov:2017kmk} (see also \cite{Kozcaz:2018usv}), setting $q$ to a root of unity effectively compactifies $C=D^2$ replacing the infinitely-generated chiral ring $H^{*,*}_{U(1)_q} (\CM_H (G,D^2))$ with a finitely-generated $H^{*,*} (X_p)$, {\it i.e.}~replacing the infinite-dimensional space $\CM_H (G,D^2)$ by a finite-dimensional space $X_p$.
Here, $X_p$ is the Coulomb branch of the resulting 3d $\CN=4$ theory, which has the property that $Z_{\text{RW}[X]} (M_3)$ equal the the invariants computed from a non-semisimple MTC associated with $\bar \CU_q (\frak{sl}_2)$ at the $2p$-th root of unity.
From the discussion in Section~\ref{sec:Verlinde}, we know that $X_p$ has $p-1$ fixed points and $\text{MTC} [X_p]$ has $S$ and $T$ matrices of size $3p-1$. For example, $X_2 = \C^2$ and, in a similar way, one can identify $X_p$ for other roots of unity. We expect $X_p$'s to be close cousins of the spaces \cite{MR2349618,MR3932780} that appear in the context of the small quantum group.
While the explicit description of $X_p$ is very desirable (and we hope to report on it in the future), it will not be necessary for what follows.
Indeed, the important point for the purpose of the present discussion is that $D^2 \times S^1$ as well as $q$ (often written together as $D^2 \times_q S^1$) represent background in the part\footnote{In 3d-3d correspondence, this is the part of the space-time where 3d $\CN=2$ theory $T[M_3]$ lives.} of the 6d space-time \eqref{fivebrane} that does not involve $M_3$. For all choices of $C$ and all values of $q$, including roots of unity, it preserves $\CN=4$ supersymmetry in the remaining three dimensions which, when topologically twisted, leads to a Rozansky-Witten theory on $M_3$ with target $X_p$.

One can also motivate Conjecture~\ref{conj:FKADO} and $\text{MTC} [X_p]$ by interpreting $\hat Z$-invariants as a non-perturbative definition of the analytically continued Chern-Simons theory with complex gauge group $G_{\C}$. In this theory, the infinite-dimensional Hilbert space $\CH (T^2)$ carries an action of the modular group $\text{MCG} (T^2) = SL(2,\Z)$ that plays an important role in surgery operations \cite{Gukov:2019mnk,Gukov:2003na}. At special values of $q = \zeta_p$ it contains a finite-dimensional subrepresentation associated with compact submanifolds in the character variety $\CM_{\text{flat}} (G_{\C}, T^2)$. For example, in the case of $G=SU(2)$, it contains five components: one semisimple and four unipotent \cite{Chun:2019mal}. Quantization of the semisimple component, often denoted $\CM_{\text{flat}} (G,C)$, gives the Verlinde representation, while incorporating the four unipotent components leads to a non-semisimple theory. In particular, the five compact cycles ($\CM_{\text{flat}} (G,C)$ and the four unipotent components) correspond to the five objects of $\text{MTC} [X_2]$ when $p=2$.

Before we present evidence to Conjecture~\ref{conj:FKADO} and test it for simple knots, a few important remarks are in order. First, it is important to stress that the standard conventions for $q$ used in $F_K (x,q)$ and in ADO invariants are related by $q \mapsto q^2$. That is why, in~\eqref{FKADOconj}, $F_K$ at $q = e^{2\pi i/p}$ corresponds to the ADO invariant for the root of unity $e^{\pi i/p}$.
Another, equivalent, explanation of this relation is that the non-semisimple category $\text{MTC} [X_p]$ associated to $\bar \CU_q (\frak{sl}_2)$ at the $2p$-th root of unity via semisimplification gives the Verlinde category at level $k=p-1$.

Another important comment is that, just like the Jones polynomial or the HOMFLY-PT polynomial, the knot invariants $F_K (x,q)$ and $\text{ADO}_p (x;K)$ can be {\it normalized} or {\it unnormalized}. Therefore, in writing a relation between them one finds four possible normalization choices, and \eqref{FKADOconj} assumes that both are normalized or both are unnormalized.
In the literature on ADO invariants, the normalized version is more popular, though the unnormalized version can also be found {\it e.g.} in \cite{MR2569561}.
On the other hand, the invariants $F_K$ are usually presented in the unnormalized form, and the normalized version is obtained by dividing by $x^{\frac{1}{2}} - x^{-\frac{1}{2}}$,
\be
F_K^{\text{norm}} (x,q) \; = \; \frac{F_K (x,q)}{x^{1/2} - x^{-1/2}}
\label{FnormF}
\ee
and similarly for the ADO invariants.
Therefore, if we wish to relate the more commonly used versions, unnormalized $F_K$ and normalized ADO, the relation \eqref{FKADOconj} should read
\be
\left. F_K (x,q) \right|_{q = \zeta_p} \; = \;
\frac{\text{ADO}_p (x/\zeta_p;K)}{\Delta_K (x^{p})} \cdot \big( x^{1/2} - x^{-1/2} \big)
\,, \qquad \zeta_p := e^{2 \pi i /p}.
\label{FKADOnorm}
\ee
As a general rule of thumb, the normalized versions of $F_K$ and ADO invariants are symmetric under $x \leftrightarrow x^{-1}$, whereas the unnormalized ones are anti-symmetric.

\begin{table}[ht]
	\centering
	\begin{tabular}{ccc}
		\hline\hline
		$p$ &  & $\phantom{\oint_{\oint_{\oint}}^{\oint^{\oint}}}
		\text{ADO}_p (x/\zeta_p;{\bf 3_1}) \; = \; F_{{\bf 3_1}} (x,q=\zeta_p) \cdot \frac{\Delta_{{\bf 3_1}} (x^p)}{x^{1/2} - x^{-1/2}}$
		\\
		\hline\hline
		$1$ & & $1$ \\[2ex]
		$2$ & & $-x-1-x^{-1}$ \\[2ex]
		$3$ & & $\zeta_3 x^2 + \zeta_3 x + (\zeta_3 - \zeta_3^{-1}) + \zeta_3 x^{-1} + \zeta_3 x^{-2}$ \\[2ex]
		$4$ & & $i x^3 + i x^2 + (1+i) x + (1 + 2i) + (1 + i) x^{-1} + i x^{-2} + i x^{-3}$ \\[1ex]
		\hline\hline
	\end{tabular}
	\caption{Normalized ADO invariants for the right-handed trefoil knot $K={\bf 3_1}$, evaluated at $x/\zeta_p$ as limits of the unnormalized 2-variable series $F_K (x,q)$.}
	\label{tab:Ftrefoil}
\end{table}

Let us illustrate this in some detail for a simple example of the trefoil knot, $K={\bf 3_1}$, and then present results of the similar computations for other knots. In all cases that we checked, Conjecture~\ref{conj:FKADO} holds. We are going to use the normalizations as in \eqref{FKADOnorm}, to simplify comparison with the literature on the ADO side as well as on the $F_K$ side.
In particular, starting with the unnormalized version of the 2-variable series $F_{{\bf 3_1}} (x,q)$ presented in \eqref{FKtrefoil} and specializing to $q=\zeta_p = -1$ that corresponds to $p=2$, we get
\be
\left. F_{{\bf 3_1}} (x,q) \right|_{q = -1}
\; = \; - \frac{x^{3/2} - x^{-3/2}}{x^2 - 1 + x^{-2}}.
\ee
As expected, the denominator of this expression is precisely the Alexander polynomial of the trefoil knot $\Delta_{{\bf 3_1}} (x) = x - 1 + x^{-1}$ evaluated at $x^2$. The numerator $x^{3/2} - x^{-3/2}$ is the unnormalized version of the ADO polynomial $\text{ADO}_2 (x;{\bf 3_1})$. In order to obtain a more familiar, normalized version we use \eqref{FKADOnorm} and divide by $x^{1/2} - x^{-1/2}$:
\be
- \frac{x^{3/2} - x^{-3/2}}{x^{1/2} - x^{-1/2}}
\; = \; - x - 1 - x^{-1}
\; = \; \text{ADO}_2 (-x;{\bf 3_1})
\; = \; \Delta_{{\bf 3_1}} (-x).
\ee
Note, the last equality is in perfect agreement with \eqref{ADOAlex}. Similarly, substituting $q = e^{2\pi i /p}$ into \eqref{FKtrefoil}, it is an easy and fun exercise to see how the infinite series collapses into a rational function of $x$ for other values of $p$. As expected, the denominator is the Alexander polynomial evaluated at $x^p$, whereas the numerator gives the $p$-th ADO polynomial. See Table~\ref{tab:Ftrefoil} for the first few values of $p$.

\begin{table}[ht]
	\centering
	\begin{tabular}{ccc}
		\hline\hline
		$p$ &  & $\phantom{\oint_{\oint_{\oint}}^{\oint^{\oint}}}
		\text{ADO}_p (x/\zeta_p;{\bf 4_1}) \; = \; F_{{\bf 4_1}} (x,q=\zeta_p) \cdot \frac{\Delta_{{\bf 4_1}} (x^p)}{x^{1/2} - x^{-1/2}}$
		\\
		\hline\hline
		$1$ & & $1$ \\[2ex]
		$2$ & & $x+3+x^{-1}$ \\[2ex]
		$3$ & & $x^2+3x+5+3x^{-1}+x^{-2}$ \\[2ex]
		$4$ & & $x^3+3x^2+6x+7+6x^{-1}+3x^{-2}+x^{-3}$\\[2ex]
		$5$ & & $x^4+3x^3+(6 + \zeta_5 + \zeta_5^{-1}) x^2 + (9 + \zeta_5 + \zeta_5^{-1}) x + 10 + \ldots$ \\[1ex]
		\hline\hline
	\end{tabular}
	\caption{Normalized ADO invariants for the figure-8 knot $K={\bf 4_1}$, evaluated at $x/\zeta_p$ as limits of the unnormalized 2-variable series $F_K (x,q)$.}
	\label{tab:Ffigure8}
\end{table}

The computations for other knots are similar, though details become progressively more involved. In Tables \ref{tab:Ffigure8} and \ref{tab:F52}, we illustrate \eqref{FKADOnorm} for the hyperbolic knots ${\bf 4_1}$ and ${\bf 5_2}$, respectively.

\begin{table}[ht]
	\centering
	\begin{tabular}{ccc}
		\hline\hline
		$p$ &  & $\phantom{\oint_{\oint_{\oint}}^{\oint^{\oint}}}
		\text{ADO}_p (x/\zeta_p;{\bf 5_2}) \; = \; F_{{\bf 5_2}} (x,q=\zeta_p) \cdot \frac{\Delta_{{\bf 5_2}} (x^p)}{x^{1/2} - x^{-1/2}}$
		\\
		\hline\hline
		$1$ & & $1$ \\[2ex]
		$2$ & & $-2x-3-2x^{-1}$ \\[2ex]
		$3$ & & $-(3 + 2\zeta_3) x^2 - (5 + 4\zeta_3)x - (7+6\zeta_3) - (5 + 4\zeta_3)x^{-1} - (3 + 2\zeta_3)x^{-2}$ \\[2ex]
		$4$ & & $-(2+2i)x^3 - (3+5i)x^2 - (4+8i)x  - (5+10i) - (4+8i)x^{-1} - \ldots$\\[1ex]
		\hline\hline
	\end{tabular}
	\caption{Normalized ADO invariants for the hyperbolic knot $K={\bf 5_2}$, evaluated at $x/\zeta_p$ as limits of the unnormalized 2-variable series $F_K (x,q)$.}
	\label{tab:F52}
\end{table}

The physical motivation for Conjecture~\ref{conj:FKADO} mentioned earlier has another interesting consequence: it predicts that the ADO invariants can be categorified. Indeed, the setup discussed here is a close cousin of the one used in the physical realization of the Heegaard Floer homology and its relation to the holomogy theory categorifying $\hat Z$-invariants~\cite{Gukov:2016gkn}.
Similarly, we expect the homology theory categorifying the ADO invariants to be infinite-dimensional, related to the homology of knot complements categorifying $F_K$ by a spectral sequence compatible with the collapse of the $q$-grading $\Z \to \Z_p$.
We also expect non-integer coefficients of ADO polynomials to emerge from the Euler characteristic of this infinite-dimensional homology by the same mechanism as non-integer values of the inverse Turaev-Milnor torsion emerge from limits of $\hat Z$-invariants~\cite{Chun:2019mal}.

\subsection*{A-polynomial and ADO polynomials}

A simple but interesting corollary of Conjecture~\ref{conj:FKADO} is that ADO polynomials should obey $q$-difference equations:
\be
\hat A^{(p)} \, \text{ADO}_p (x) \; = \; 0.
\label{AAADO}
\ee
Indeed, quantization of character varieties in Chern-Simons theory leads to the statement \cite{Gukov:2003na} that the partition function on the knot complements should be annihilated by the quantum A-polynomial, $\hat A \, Z \left( S^3 \setminus K \right) =0$, where $\hat A = \hat A (\hat x, \hat y; K, q)$ can be explicitly computed from the all-order perturbative expansion \cite{Dimofte:2009yn}. Similar arguments imply that $F_K (x,q)$ also obeys the same $q$-difference equation \cite{Gukov:2019mnk}:
\be
\hat A (\hat x , \hat y; K, q) \, F_K (x,q) \; = \; 0
\ee
which can be also deduced directly from the fivebrane setup \eqref{fivebrane}. Taking the limit $q \to \zeta_p$ in both $\hat A (\hat x , \hat y; K, q)$ and in $F_K (x,q)$, we obtain \eqref{AAADO} where
\be
\hat y \hat x \; = \; \zeta_p \, \hat x \hat y
\label{xycomm}
\ee
and $\hat A^{(p)} = \hat A^{(p)} (\hat x , \hat y; K)$ is obtained from $\hat A (\hat x, \hat y; K, q = \zeta_p)$ by conjugating with the ``normalization factor'' in \eqref{FKADOnorm},
\be
\hat A^{(p)} (\hat x , \hat y; K)
\; = \;
\frac{\Delta_K (x^{p})}{( x^{1/2} - x^{-1/2} )}
\left. \hat A (\hat x , \hat y; K, q) \right|_{q = \zeta_p} 
\frac{( x^{1/2} - x^{-1/2} )}{\Delta_K (x^{p})}.
\ee
The explicit form of $\hat A^{(p)} (\hat x , \hat y; K)$ for the trefoil knot $K={\bf 3_1}$ and for the figure-8 knot $K={\bf 4_1}$ can be found in Tables~\ref{tab:A31} and \ref{tab:A41}, respectively.

\begin{table}[ht]
	\centering
	\begin{tabular}{ccc}
		\hline\hline
		$p$ &  & $\phantom{\oint_{\oint_{\oint}}^{\oint^{\oint}}}
		\hat A^{(p)} (x , y; {\bf 3_1})$
		\\
		\hline\hline
		$2$ & & $\phantom{\oint_{\oint_{\oint}}^{\oint^{\oint}}} 1 + (1-2x+2x^2-x^3) y - x^3 y^2$ \\[2ex]
		$3$ & & $1 - \frac{1 - \zeta_3^{-1} x - x^2 + \zeta_3 x^4 + \zeta_3 x^5 - x^6}{\zeta_3 (1 - \zeta_3 x)(1 - \zeta_3^{-1} x^2)} y - \frac{x^3 (1-x) (1-x^2)}{(1 - \zeta_3 x)(1 - \zeta_3^{-1} x^2)} y^2$ \\[2ex]
		$4$ & & $(1-ix)(1-ix^2) + i (1+x+ix^2 - ix^4 - x^5 -x^6) y + (1+ix)(1+ix^2) y^2$\\[1ex]
		\hline\hline
	\end{tabular}
	\caption{Annihilators (quantum A-polynomials) for the ADO invariants of the trefoil knot $K={\bf 3_1}$ for the first few values of $p$. Even though $x$ and $y$ here obey $y x = \zeta_p xy$, to avoid clutter we write $x$ and $y$ in place of $\hat x$ and $\hat y$.}
	\label{tab:A31}
\end{table}

\begin{table}[ht]
	\centering
	\begin{tabular}{ccc}
		\hline\hline
		$p$ &  & $\phantom{\oint_{\oint_{\oint}}^{\oint^{\oint}}}
		\hat A^{(p)} (x , y; {\bf 4_1})$
		\\
		\hline\hline
		$2$ & & $\phantom{\oint_{\oint_{\oint}}^{\oint^{\oint}}} 1 - (x^{-2}-3x^{-1}+3-3x+x^2) y- (x^{-2}+3x^{-1}+3+3x+x^2) y^2 + y^3$ \\[2ex]
		$3$ & & $x^2 (x+1) (1-\zeta_3 x) (1-\zeta_3 x^2)$ \\[1ex]
		& & $-\zeta_3^2
		(1-\zeta_3^{-1} x)
		(1 + \zeta_3^{-1} x)
		(1 - \zeta_3 x^2)
		(\zeta_3^{-1} x^4 - \zeta_3^{-1} (1+2 \zeta_3 x) x^2$ \\[1ex]
		& & $ + (\zeta_3^{-1} x^2 + \zeta_3 x + 1) x - \zeta_3^{-1} (2+ \zeta_3 x) x + 1) y$ \\[1ex]
		& & $+\zeta_3
		(1-x^2)^2
		(x^4+\zeta_3 x^3 - 2 x^3 - \zeta_3 x^2 -\zeta_3^{-1} x^2 + x^2 + \zeta_3^{-1} x - 2x + 1) y^2$ \\[1ex]
		& & $-\zeta_3 x^2
		(1-\zeta_3 x)
		(1+\zeta_3^{-1} x)
		(1-x^2) y^3$ \\[2ex]
		$4$ & & $1+\frac{1-x \left(x^5+(2+i) x^4-2 i x^2+(-2+i)\right)}{x^2 (x+i)^2} y-\frac{x^6-(1+2 i) x^5-2 x^3-(1-2 i) x+1}{i x^2 (x+i)^2}y^2 + \frac{i (x-1)^2}{(x+i)^2} y^3$\\[1ex]
		\hline\hline
	\end{tabular}
	\caption{Annihilators (quantum A-polynomials) for the ADO invariants of the figure-8 knot $K={\bf 4_1}$ for the first few values of $p$. Even though $x$ and $y$ here obey $y x = \zeta_p xy$, to avoid clutter we write $x$ and $y$ in place of $\hat x$ and $\hat y$.}
	\label{tab:A41}
\end{table}

\subsection*{New knot invariants and higher-rank quantum groups}

To the best of our knowledge, the higher-rank analogues of the ADO invariants have not been worked out.\footnote{Part of the difficulty is that the detailed structure of projective modules was understood only recently using the tools of geometric representation theory~\cite{MR2349618,MR3932780}, and mostly in the context of the small quantum group $\CU_q^{\text{small}} (\frak{sl}_N)$ at \emph{odd} root of unity. The restricted quantum group $\bar \CU_q (\frak{sl}_2)$ at \emph{even} root of unity is defined in the same manner, but must be discussed with care. See \cite{2018arXiv181202277N}.} This gives us an opportunity to make new predictions for what they should be, based on the proposed relations \eqref{theweb}.

In the case of $\frak{g} = \frak{sl}_N$, the invariants \eqref{FKcomplement} depend on $q$ and $x \in \mathbb{T}_{\C}$, where $\mathbb{T}_{\C}$ is the maximal torus\footnote{The argument is the same as in the generalized volume conjecture \cite{Gukov:2003na}, and the resulting invariants can often be formulated in terms of a ``state integral'' model {\it a la} \cite{Dimofte:2009yn} that involves states labeled by $\mathbb{T}_{\C}$-valued variables \cite{Gukov:2016gkn,Chung:2018rea,Park:2019xey}.} of $G_{\C} = SL(N,\C)$.
Following the same arguments that led us to Conjecture~\ref{conj:FKADO}, we expect $\left. F_K^{SU(N)} (x,q) \right|_{q = e^{2\pi i/p}}$ to be a rational function of $x$, with the denominator $\prod_{i < j} \Delta_K ((x_i/x_j)^{p})$ given by a product (over $i,j=1, \ldots N$) of several copies of the Alexander polynomial. We are more interested, however, in the numerator of this rational function. Since in the higher-rank version of \eqref{FnormF} the factor $x^{1/2} - x^{-1/2}$ is replaced \cite{Park:2019xey} by a product $\prod_{\alpha \in \Delta^+} (x^{\alpha/2} - x^{- \alpha/2})$ over positive roots of $\frak{g}$, following \eqref{FKADOnorm} we define the higher-rank analogues of the {\it normalized} ADO invariants to be
\be
P_K^{SU(N)} (x,p) \; = \; \left. F_K^{SU(N)} (x,q) \right|_{q = e^{2\pi i/p}} \cdot \prod_{\alpha \in \Delta^+} \frac{\Delta_K (x^{p\alpha})}{x^{\alpha/2} - x^{- \alpha/2}}.
\label{PKdef}
\ee
We expect $P_K^{SU(N)} (x,p)$ to be Laurent polynomials in $x$ for all $K$, for all $p$, and for all $N$.

Using the explicit computations of $F_K^{SU(N)} (x,q)$ in the recent work \cite{Park:2019xey}, one can compute the specializations \eqref{PKdef} for various knots. It would be interesting to reproduce these polynomials $P_K^{SU(N)} (x,p)$ by other methods, especially rooted in the representation theory of $\CU_q (\frak{sl}_N)$. In Table~\ref{tab:FSU3} we present the first few polynomials for the right-handed trefoil knot $K = {\bf 3_1}$. It is curious to note that $P_{{\bf 3_1}}^{SU(3)} (x,2) = \Delta_{{\bf 3_1}} (-x_1) \cdot \Delta_{{\bf 3_1}} (-x_2) \cdot \Delta_{{\bf 3_1}} (-x_1 x_2) + 2$.

\begin{table}[ht]
	\centering
	\begin{tabular}{ccc}
		\hline\hline
		$p$ &  & $\phantom{\oint_{\oint_{\oint}}^{\oint^{\oint}}}
		P_{{\bf 3_1}}^{SU(3)} (x,p)$
		\\
		\hline\hline
		$1$ & & $1$ \\[2ex]
		$2$ & & $1-2x_1-2x_2-2x_1^{-1}x_2^{-1}-2x_1x_2-2x_1^{-1}-2x_2^{-1}
		- x_1^2x_2-x_2x_1^{-1}-x_1^{-1}x_2^{-2}$ \\[1ex]
		& & $-x_1x_2^2-x_1x_2^{-1}-x_1^{-2}x_2^{-1}
		-x_1^2-x_2^2-x_1^{-2}x_2^{-2}-x_1^2x_2^2-x_1^{-2}-x_2^{-2}$ \\[2ex]		
		$3$ & & $-1-7\zeta_3 -4\zeta_3(x_1+x_2+x_1^{-1}x_2^{-1}+x_1 x_2+x_1^{-1}+x_2^{-1}) + x_1^4 +x_1^4 x_2^4 +x_2^4 +x_1^{-4}$ \\[1ex]
		& & $+x_1^{-4}x_2^{-4}+x_2^{-4}+(2-2\zeta_3)(x_1^2 x_2+x_1^{-1}x_2+x_1^{-1}x_2^{-2}+x_1 x_2^2+x_1 x_2^{-1}+x_1^{-2}x_2^{-1})$\\[1ex]
		& & $+(3-2\zeta_3)(x_1^2+x_2^2+x_1^{-2}x_2^{-2}+x_1^2 x_2^2+x_1^{-2}+x_2^{-2}) +x_1^{-1}x_2^3 +x_1^{-3}x_2  +x_1^{-4}x_2^{-1}$ \\[1ex]
		& & $+(2-\zeta_3)(x_1^3 x_2 +x_1^3 x_2^2 +x_1^2 x_2^3 +x_1 x_2^3 +x_1^{-1}x_2^2 +x_1^{-2}x_2 +x_1^{-3}x_2^{-1} +x_1^{-3}x_2^{-2}$\\[1ex]
		& & $+x_1^{-2}x_2^{-3} +x_1^{-1}x_2^{-3} +x_1 x_2^{-2} +x_1^2 x_2^{-1}) +x_1^4 x_2 +x_1^4 x_2^3 +x_1^3 x_2^4  +x_1 x_2^4+x_1^{-4}x_2^{-3}$ \\[1ex]
		& & $+2(x_1^3 +x_1^3x_2^3 +x_2^3 +x_1^{-3} +x_1^{-3}x_2^{-3} +x_2^{-3}) +x_1^{-3}x_2^{-4} +x_1^{-1}x_2^{-4} +x_1 x_2^{-3}$ \\[1ex]
		& & $+(1-\zeta_3)(x_1^4 x_2^2 +x_1^2 x_2^4 +x_1^{-2}x_2^2 +x_1^{-4}x_2^{-2} +x_1^{-2}x_2^{-4} +x_1^2 x_2^{-2}) +x_1^3 x_2^{-1}$ \\[1ex]
		\hline\hline
	\end{tabular}
 	\caption{Analogues of the normalized ADO invariants for the right-handed trefoil knot and $G=SU(3)$, produced from the unnormalized 2-variable series $F_{{\bf 3_1}} (x,q)$ via~\eqref{PKdef}.}
	\label{tab:FSU3}
\end{table}

\subsection*{3-manifolds}

Finally, we note that Conjecture~\ref{conj:FKADO} has another obvious generalization.
Indeed, just like Conjecture~\ref{conj:ZRW}, which is expected to hold for link complements as well as closed 3-manifolds without boundary, the proposed relations \eqref{FKADOconj} and \eqref{FKADOnorm} have an obvious analogue for closed 3-manifolds.
Testing such ``Conjecture~\ref{conj:FKADO} for closed 3-manifolds'' is likely to require a careful analysis of normalizations and other conventions, as we did in going from the general statement \eqref{FKADOconj} to its version \eqref{FKADOnorm} written in the most commonly used set of conventions.

\acknowledgments{It is pleasure to thank J{\o}rgen Andersen, Francesco Costantino, Pavel Etingof, Boris Feigin, Igor Frenkel, Azat Gainutdinov, Amihay Hanany, Anna Lachowska, Ciprian Manolescu, Jun Murakami, Mark Penney, Lev Rozansky, and Shing-Tung Yau for illuminating discussions and comments.
The work of S.G.\ is supported by the U.S. Department of Energy, Office of Science, Office of High Energy Physics, under Award No. DE-SC0011632, and by the National Science Foundation under Grant No. NSF DMS 1664240.
The work of P.-S. H. is supported by the U.S. Department of Energy, Office of Science, Office of High Energy Physics, under Award Number DE-SC0011632, and by the Simons Foundation through the Simons Investigator Award.
The work of H.N.\ is supported in part by the World Premier International Research Center Initiative (WPI Initiative), MEXT, Japan, and by JSPS Grant Number 16H06335, 19K21828. 
The work of S.P.\ is supported by Kwanjeong Educational Foundation. The work of D.P.\ is supported by the Center for Mathematical Sciences and Applications at Harvard University. N.S. gratefully acknowledges the support of the Dominic Orr Graduate Fellowship at Caltech.}


\appendix

\section{Topological twists of 3d $\CN=4$ theories}
\label{app:3dtwists}

Three-dimensional $\CN=4$ supersymmetry algebra allows a vector multiplet, that contains 1-form gauge field, and two types of matter multiplets that can be used to parametrize Higgs and Coulomb branches of 3d $\CN=4$ gauge theories. These two types of matter multiplets are sometimes called hypermultilets and twisted hypermultiplets (where ``twisted'' has nothing to do with the topological twist), see {\it e.g.} \cite{Kapustin:1999ha}.

If, following \cite{Rozansky:1996bq} (and also \cite{Blau:2000iy}), we denote the R-symmetry of 3d $\CN=4$ theory by $SU(2)_R \times SU(2)_N$, then the two topological twists of 3d $\CN=4$ theories can be characterized by their action on fields in the supermultiplets. One topological twist, when applied to 3d $\CN=4$ gauge theory, localizes on solutions to 3d Seiberg-Witten equations and their generalizations:
$$
\text{Vector}: \qquad
\begin{array}{rcl}
& SU(2)_E \times SU(2)_R \times SU(2)_N & \to \quad SU(2)_E' \times SU(2)_N  \\
\text{bosons:} & ({\bf 3}, {\bf 1}, {\bf 1}) \oplus ({\bf 1}, {\bf 1}, {\bf 3}) & \to \quad
({\bf 3}, {\bf 1}) \oplus ({\bf 1}, {\bf 3}) \\
\text{fermions:} & ({\bf 2}, {\bf 2}, {\bf 2} ) & \to \quad ({\bf 3}, {\bf 2}) \oplus ({\bf 1}, {\bf 2})
\end{array}
$$
$$
\text{Hyper}: \qquad
\begin{array}{rcl}
& SU(2)_E \times SU(2)_R \times SU(2)_N & \to \quad SU(2)_E' \times SU(2)_N  \\
\text{bosons:} & 2 ({\bf 1}, {\bf 2}, {\bf 1}) & \to \quad
2 ({\bf 2}, {\bf 1}) \\
\text{fermions:} & ({\bf 2}, {\bf 1}, {\bf 2} ) & \to \quad ({\bf 2}, {\bf 2})
\end{array}
$$
$$
{\text{Twisted} \atop \text{hyper}:} \qquad
\begin{array}{rcl}
& SU(2)_E \times SU(2)_R \times SU(2)_N & \to \quad SU(2)_E' \times SU(2)_N  \\
\text{bosons:} & 2 ({\bf 1}, {\bf 1}, {\bf 2}) & \to \quad
2 ({\bf 1}, {\bf 2}) \\
\text{fermions:} & ({\bf 2}, {\bf 2}, {\bf 1} ) & \to \quad ({\bf 3}, {\bf 1}) \oplus ({\bf 1}, {\bf 1})
\end{array}
$$
Here, $SU(2)_E'$ is the diagonal subgroup of $SU(2)_E \times SU(2)_R$.
The second topological twist, when applied to gauge theory, localizes on complex flat connections \cite{Blau:1996bx} and acts on the fields as
$$
\text{Vector}: \qquad
\begin{array}{rcl}
& SU(2)_E \times SU(2)_R \times SU(2)_N & \to \quad SU(2)_E' \times SU(2)_R  \\
\text{bosons:} & ({\bf 3}, {\bf 1}, {\bf 1}) \oplus ({\bf 1}, {\bf 1}, {\bf 3}) & \to \quad
({\bf 3}, {\bf 1}) \oplus ({\bf 3}, {\bf 1}) \\
\text{fermions:} & ({\bf 2}, {\bf 2}, {\bf 2} ) & \to \quad ({\bf 3}, {\bf 2}) \oplus ({\bf 1}, {\bf 2})
\end{array}
$$
$$
\text{Hyper}: \qquad
\begin{array}{rcl}
& SU(2)_E \times SU(2)_R \times SU(2)_N & \to \quad SU(2)_E' \times SU(2)_R  \\
\text{bosons:} & 2 ({\bf 1}, {\bf 2}, {\bf 1}) & \to \quad
2 ({\bf 1}, {\bf 2}) \\
\text{fermions:} & ({\bf 2}, {\bf 1}, {\bf 2} ) & \to \quad ({\bf 3}, {\bf 1}) \oplus ({\bf 1}, {\bf 1})
\end{array}
$$
where $SU(2)_E'$ is the diagonal subgroup of $SU(2)_E \times SU(2)_N$.
Note, under these two twists the role of hyper and twisted hyper-multiplets is interchanged, {\it i.e.} the first twist applied to ordinary hypermultiplets leads to the same topological theory as the second twist applied to twisted hypermultiplets, and vice versa.\footnote{One can also label these two twists by ``Coulomb'' or ``Higgs'' (or ``C'' and ``H'' for short) based on the part of the R-symmetry which is either used in the topological twist or remains preserved by the twist, {\it cf.}~\cite{Bullimore:2018jlp}.}

\section{Quantum equivariant K-theory of \texorpdfstring{$T^*\mathbb{C}\mathbf{P}^1$}{T^*CP^1}}
\label{app:ktheory}

$QK_T(T^* \mathbb{C}\mathbf{P}^1)$ corresponds to the algebra of line operators and is generated by two elements $I$ and $W$, corresponding to Wilson loops of charge 0 and charge 1. These Wilson operators correspond to the powers of tautological bundle $V$. There is a natural pairing on a classical K-theory
\be
\eta_{\mathcal{A} \mathcal{B}}  = \chi(X, \mathcal{A} \otimes \mathcal{B} \otimes \mathcal{K}_X^{-1/2})
\ee
which is not deformed, and which can be easily computed using localization. Here $\mathcal{K}^{1/2}$ is ``the half'' of the canonical bundle in the natural polarization on $X$, which is necessary in order to identify the trivial Wilson line $I$ and the structure sheaf. In our case
\be
\eta_{I I} = \frac{\sqrt{x_1/x_2}+\sqrt{x_2/x_1}}{(1-t^{-1} x_2/x_1)(1-t^{-1} x_1/x_2)},
\ee
\be
\frac{\eta_{W I}}{\eta_{I I}} = \frac{1+t^{-1}}{x_1^{-1}+x_2^{-1}},
\ee
and
\be
\frac{\eta_{W W}}{\eta_{I I}} = t^{-1} x_1 x_2.
\ee
In order to find quantum K-theory ring we can use recursion relations for expectation values of line operators on an infinite plane using Bethe ansatz equations for vacua. If we set $\langle I \rangle_{\mathbb{R}^2 \times S^1} = 1$ then $\langle W \rangle_{\mathbb{R}^2 \times S^1} = \frac12 \sum_{vac} z_{vac}$ and we can get the ring coefficients $C^{A}_{B C}$
\be
I \cdot I = I
\ee
\be
I \cdot W = W
\ee
\be
W \cdot W = \left( \frac{\left(x_1+x_2\right) (1-q t)}{1-q t^2} \right) W + \left( -\frac{(1-q) x_1 x_2}{1-q t^2} \right) I
\ee
where $z_{vac}$ are solutions of Bethe ansatz equation
\be
q \prod_{i=1,2} \left( \frac{1-t z/x_i}{1- z /x_i} \right)=1.
\ee
These equations are obtained by requiring that an expectation value $\langle \cdot \rangle_{\mathbb{R}^2 \times S^1}$ with an arbitrary operator is the same. One can check that these ring coefficients are consistent with the metric $\eta_{B C} = C_{I B C} =\eta_{I A} C^{A}_{BC}$.

Now one can use this ring to compute partitions functions on $\Sigma \times S^1$. Note that normalization of such partition functions is different from the normalization above. More precisely here it is fixed by $Z(S^2 \times S^1) = \eta_{II}$.

\bibliography{MTCrefs}
\bibliographystyle{JHEP}
\end{document}